\documentclass[utf8]{FrontiersinHarvard}
\usepackage{url,hyperref,lineno,microtype,subcaption}
\usepackage{booktabs}
\usepackage{cleveref}
\usepackage{float}
\usepackage[onehalfspacing]{setspace}

\usepackage{tikz}
\def\checkmark{\tikz\fill[scale=0.4](0,.35) -- (.25,0) -- (1,.7) -- (.25,.15) -- cycle;}

\def\keyFont{\fontsize{8}{11}\helveticabold }
\def\firstAuthorLast{Hsiao {et~al.}} 
\def\Authors{Cheng-Hsi Hsiao\,$^{1,*}$, Krishna Kumar\,$^{1}$ and Ellen Rathje\,$^{1}$}
\def\Address{
$^{1}$Maseeh Department of Civil, Architectural and Environmental Engineering, University of Texas at Austin, Austin , TX , United States  }
\def\corrAuthor{Cheng-Hsi Hsiao}

\def\corrEmail{chhsiao@utexas.edu}

\begin{document}
\onecolumn

\title[XAI Lateral Spreading]{Explainable AI models for predicting liquefaction-induced lateral spreading} 

\author[\firstAuthorLast ]{\Authors} 
\address{} 
\correspondance{} 

\extraAuth{}

\maketitle

\begin{abstract}

\section{}
Earthquake-induced liquefaction can cause substantial lateral spreading, posing threats to infrastructure. Machine learning (ML) can improve lateral spreading prediction models by capturing complex soil characteristics and site conditions. However, the ``black box" nature of ML models can hinder their adoption in critical decision-making. This study addresses this limitation by using SHapley Additive exPlanations (SHAP) to interpret an eXtreme Gradient Boosting (XGB) model for lateral spreading prediction, trained on data from the 2011 Christchurch Earthquake. SHAP analysis reveals the factors driving the model's predictions, enhancing transparency and allowing for comparison with established engineering knowledge. The results demonstrate that the XGB model successfully identifies the importance of soil characteristics derived from Cone Penetration Test (CPT) data in predicting lateral spreading, validating its alignment with domain understanding. This work highlights the value of explainable machine learning for reliable and informed decision-making in geotechnical engineering and hazard assessment.

\tiny
 \keyFont{ \section{Keywords:} Explainable AI, SHAP, XGBoost, Lateral spreading, 2011 Christchurch earthquake} 
\end{abstract}

\section{Introduction}

Lateral spreading, characterized by the horizontal movement of liquefied soil triggered by earthquakes, can cause severe damage to infrastructure and present significant hazards in urban environments. The movement results from rapid soil transition from a solid to a liquid state during earthquake excitation, influenced by the initial state of soil and amplitude of shaking. While a site might undergo liquefaction, it will only experience lateral spreading if the topography facilitates it. Given the complex factors involved, advanced, data-driven approaches are essential for predicting the extent and impact of lateral spreading. In recent years, machine learning (ML) applications have shown promising results in modeling lateral spreading~\citep{demir-2022,kaya-2016,durante-2021}.

Although ML models extract relationships between the various influencing factors and failure occurrence, they pose interpretability and trustworthiness challenges~\citep{miller-2019,doshivelez-2017}. The term ``black box" refers to the opaque nature of these ML models, where the internal logic of their decision-making processes remains hidden. This opacity can have significant consequences, primarily when using ML models for policy decisions related to natural hazards, where inaccurate predictions could lead to substantial damage. Furthermore, ML models generally excel in interpolation, meaning they perform well when predicting scenarios similar to their training data. However, they may struggle with extrapolation, failing to accurately predict outcomes in new, unseen conditions, posing challenges in their generalizability and application to sites outside their training regime.

The complexity of lateral spreading exacerbates these challenges. Factors such as site geometry, soil type, and loading conditions influence lateral spreading, making predictions more complex. In ML models that do not consider the physics of these processes, there is a risk of including redundant or excessive features, termed feature coherence. These features could lead to model overfitting, where the model memorizes behavior rather than learning the general trend, limiting its applicability to different datasets~\citep{paris-2004,xue-2019}. Excessive coherence among features indicates that multiple inputs might overlap in their influence, causing redundancy within the model's decision-making framework and further complicating the interpretation and reliability of the model's predictions.

To enhance the transparency and interpretability of ML models, researchers have developed a variety of explainable AI (XAI) methodologies~\citep{arrieta-2020,linardatos-2020}. These approaches fall into two broad categories: model-agnostic methods and model-specific methods. Model-agnostic methods, such as Local Interpretable Model-agnostic Explanations (LIME;~\citealp{ribeiro-2016}) and SHapley Additive exPlanations (SHAP;~\citealp{lundberg-2017}), provide versatile techniques applicable across different types of ML models. These methods generate explanations by observing how the model output changes as it perturbs the input data. This process allows model-agnostic explainers to approximate how each feature influences the model's output, offering insights into its behavior without requiring access to its internal workings, making them universally applicable to any ML model. Conversely, model-specific methods, like Layer-wise Relevance Propagation (LRP;~\citealp{bach-2015}), are tailored to specific model architectures. Such XAI techniques play a pivotal role in demystifying the decision-making process of machine learning models, highlighting influential features, and facilitating the incorporation of domain knowledge for model improvement.

In this study, we utilize a model-agnostic SHAP approach to delve into the decision-making process of an XGB~\citep{Chen_2016} model designed to predict the occurrence of lateral spreading. Our analysis examines the 2011 Christchurch earthquake dataset~\citep{durante-2022}. By employing SHAP, we aim to improve the interpretability of our developed model by exploring local and global explanations for each feature. SHAP allows us to uncover complex relationships between input variables and model predictions. Notably, SHAP helps identify instances where the impact of specific features goes against established physics principles or domain knowledge. This approach contributes to a deeper understanding of lateral spreading prediction and enhances the accessibility of complex ML models for researchers and practitioners. 

The 2011 Christchurch Earthquake was a devastating event characterized by severe liquefaction and lateral spreading~\citet{cubrinovski-2011}. These phenomena, induced by the earthquake ground shaking, led to significant horizontal movement of saturated soil, causing widespread damage to infrastructure and posing substantial risks to urban areas.

\Cref{fig:displ-map} illustrates the horizontal displacement amplitudes measured using remote sensing techniques, as documented by~\citet{rathje-2017}. The map categorizes the lateral displacements into three zones: green signifies 0 to 0.3 meters, orange indicates 0.3 to 0.5 meters, and red represents displacements exceeding 0.5 meters. Our study leverages the dataset developed by~\citet{durante-2022} on the~\citet{rathje-2017} measurements to predict lateral spreading. This dataset encompasses 6,505 locations with displacement data and co-located CPT profiles. We employ a displacement threshold of 0.3 meters, as depicted by the orange and red regions in~\Cref{fig:displ-map}, to distinguish between the occurrence and non-occurrence of lateral spreading. This threshold is based on research by~\citet{rathje-2017}, which demonstrated that displacements below this value could not be statistically distinguished from areas with zero movement.~\Cref{fig:ls-map} shows the spatial distribution of the Yes/No lateral spreading data based on the threshold including 2752 Yes data (red triangle; 42.31\%) and 3753 No data (green circle; 57.69\%).
\begin{subfigure}
\setcounter{figure}{1}
\setcounter{subfigure}{0}
    \centering
    \begin{minipage}[b]{0.6\textwidth}
        \centering
        \includegraphics[width=\linewidth]{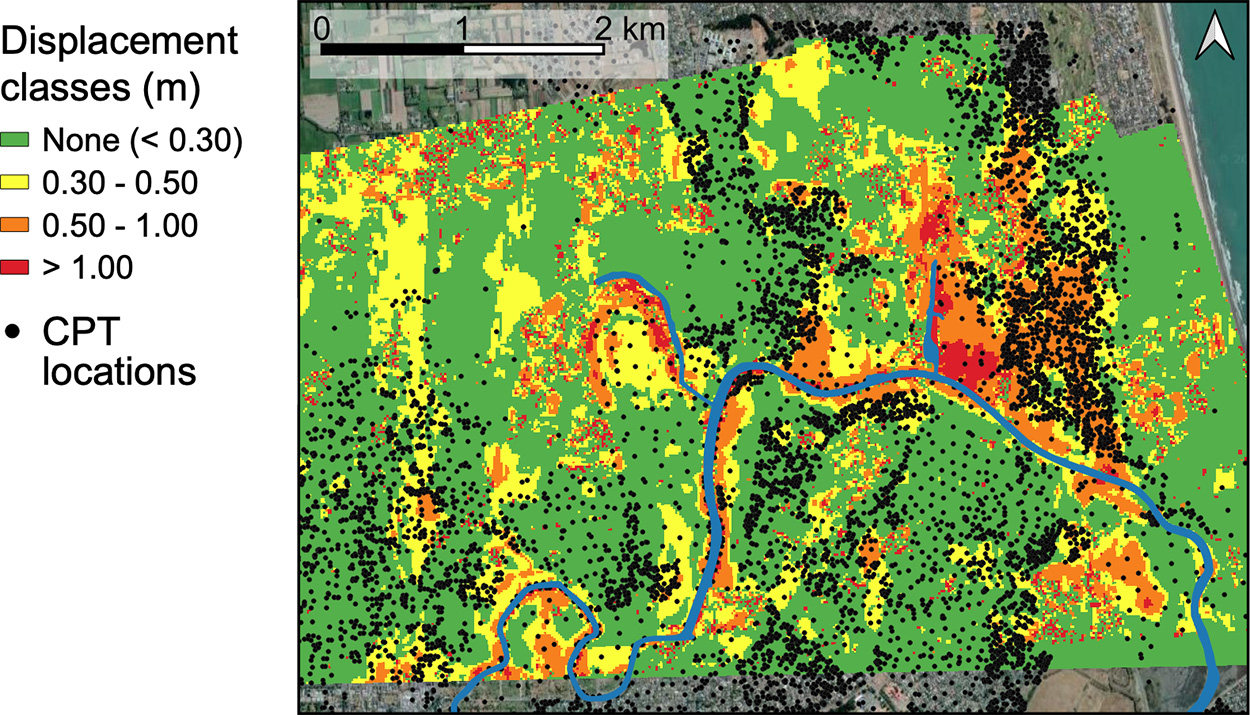}
        \caption{Observed horizontal displacement~\citep{rathje-2017}}
        \label{fig:displ-map}
    \end{minipage}
\setcounter{figure}{1}
\setcounter{subfigure}{1}
    \begin{minipage}[b]{0.475\textwidth}
        \centering
        \includegraphics[width=\linewidth]{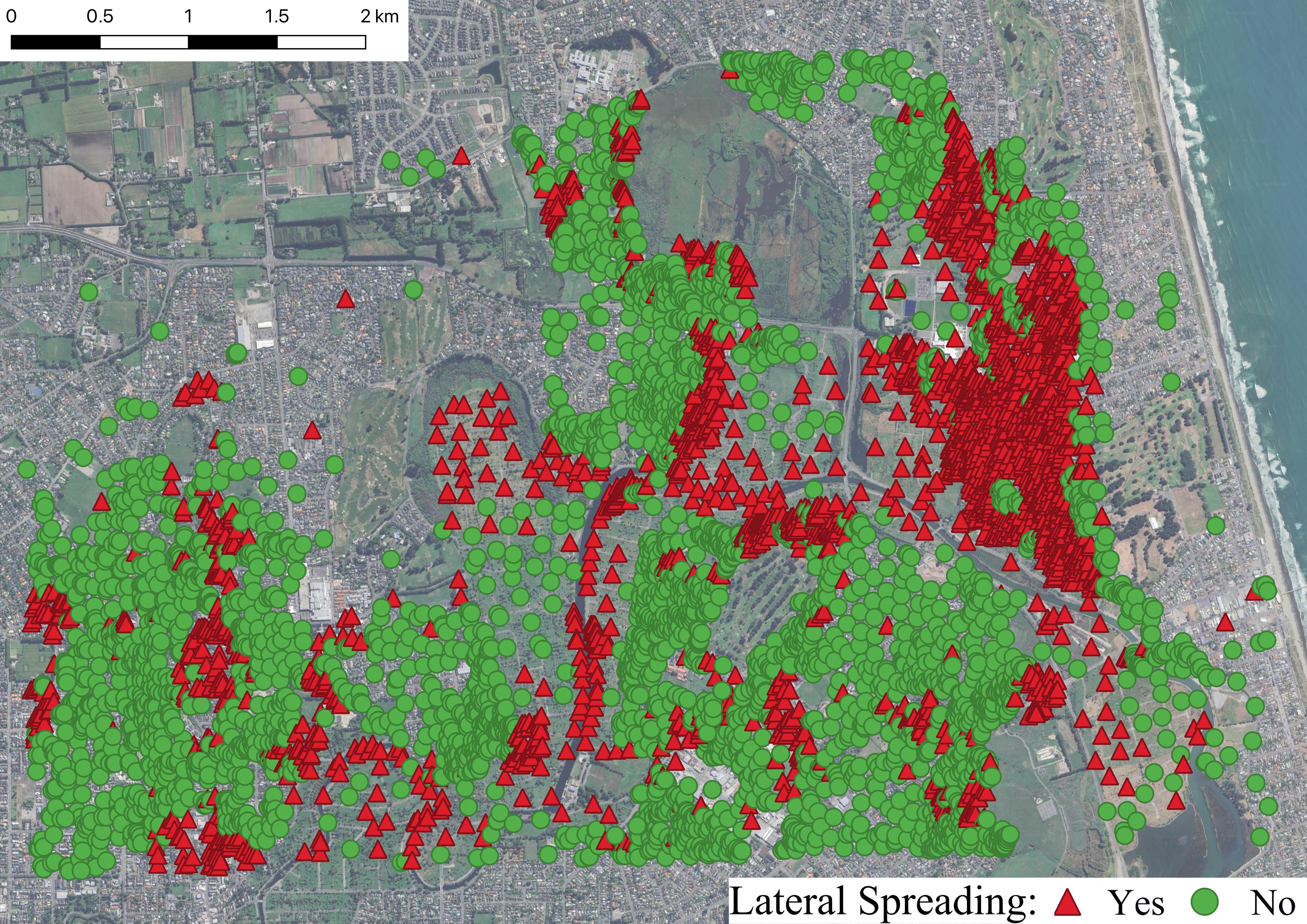}
        \caption{Lateral spreading occurrence~\citep{durante-2022}}
        \label{fig:ls-map}
    \end{minipage}
\setcounter{figure}{1}
\setcounter{subfigure}{2}
    \begin{minipage}[b]{0.475\textwidth}
        \centering
        \includegraphics[width=\linewidth]{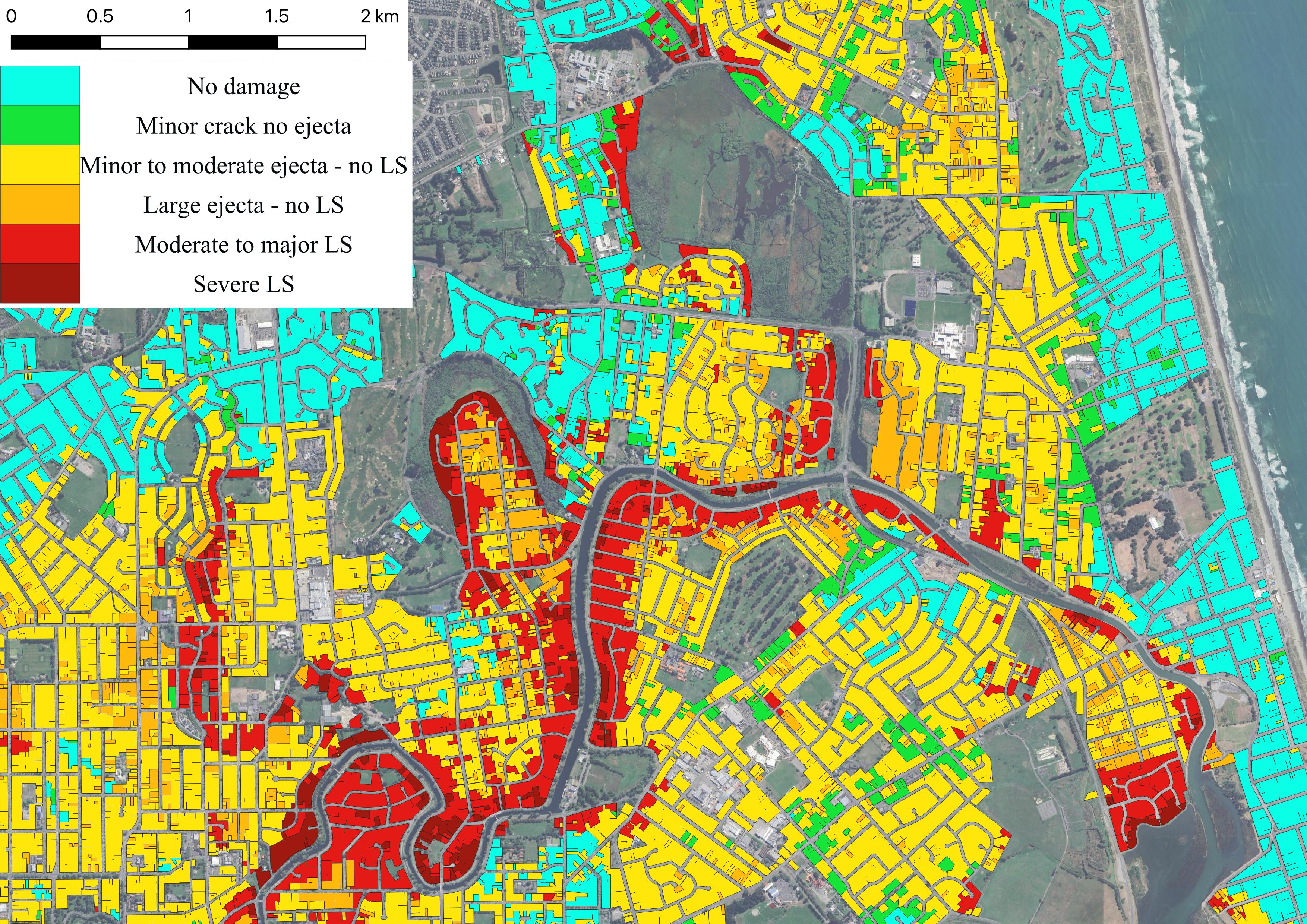}
        \caption{Observed liquefaction-related damage (NZGD)}
        \label{fig:nzgd-map}
    \end{minipage}
\setcounter{figure}{1}
\setcounter{subfigure}{-1}
    \caption{Lateral spreading occurrence map of 2011 Christchurch earthquake. \textbf{(A)} Lateral spreading displacement from remote sensing data (data from~\citealp{rathje-2017}).\textbf{(B)} Spatial distribution of lateral spreading occurrence (data from~\citealp{durante-2022}).\textbf{(C)} Observed liquefaction-related damage (data from~\citealp{nzgd-2013})}
    \label{fig:christchurch-map}
\end{subfigure}

Similar to~\cite{durante-2021}, we develop an ML model to predict lateral spreading based on five features: the distance between the site and the nearest river location (L), ground slope, ground elevation, groundwater depth (GWD), and peak ground acceleration (PGA). Our paper aims to develop a comprehensive explanation of the ML model for lateral spreading occurrence using the SHAP analysis.

\Cref{fig:data_distrib} shows the distributions of these predictive features. Each histogram is accompanied by a box diagram summarizing the feature distributions, with outliers identified as red dots using the interquartile range (IQR) method. Specifically, observations exceeding 1.5 times the IQR below the first quartile (Q1) or above the third quartile (Q3) are flagged as outliers, with the IQR representing the difference between Q1 and Q3.

We notice that groundwater depth (GWD) and slope data (\Cref{fig:gwd_distrib,fig:slope_distrib}) vary significantly in the region of interest. Specifically, we have identified 117 data points with GWD values exceeding the median value of 3.70 meters and 500 data points with slope median values exceeding 2.84\%. While these values deviate significantly from the median value, they are characteristics of the sites. Hence, we have opted to retain them in our analysis. Our rationale is to enable our trained model to learn from a broader range of feature values, enhancing its ability to generalize and make accurate predictions across various scenarios.

In this study, we conducted a comprehensive SHAP analysis to gain insights into the behavior of the XGB models developed for lateral spreading prediction. We introduce the XGB and SHAP approaches in the Methodology section. The Model Development section outlines the model configuration, training procedures, and overall performance metrics. Finally, in the Results section, we delve into the details of SHAP's local and global explanations for our trained XGB models. Through this SHAP analysis, we unveiled key insights derived from our models and assessed the significance of the CPT features in our predictive models. 
\begin{subfigure}
\setcounter{figure}{2}
\setcounter{subfigure}{0}
    \centering
    \begin{minipage}[b]{0.3\textwidth}
        \includegraphics[width=\linewidth]{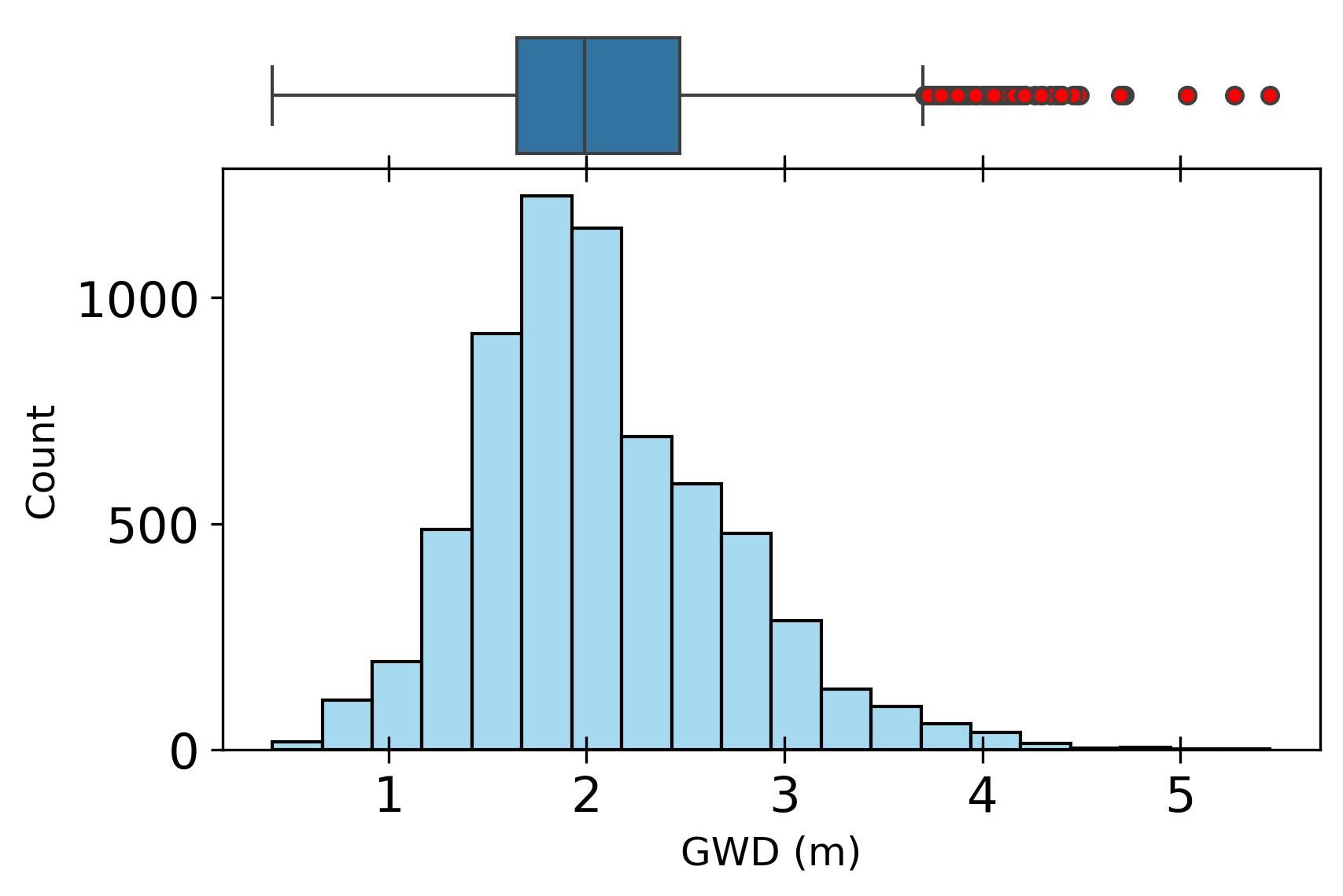}
        \caption{GWD}
        \label{fig:gwd_distrib}
    \end{minipage}  
\setcounter{figure}{2}
\setcounter{subfigure}{1}
    \centering
    \begin{minipage}[b]{0.3\textwidth}
        \includegraphics[width=\linewidth]{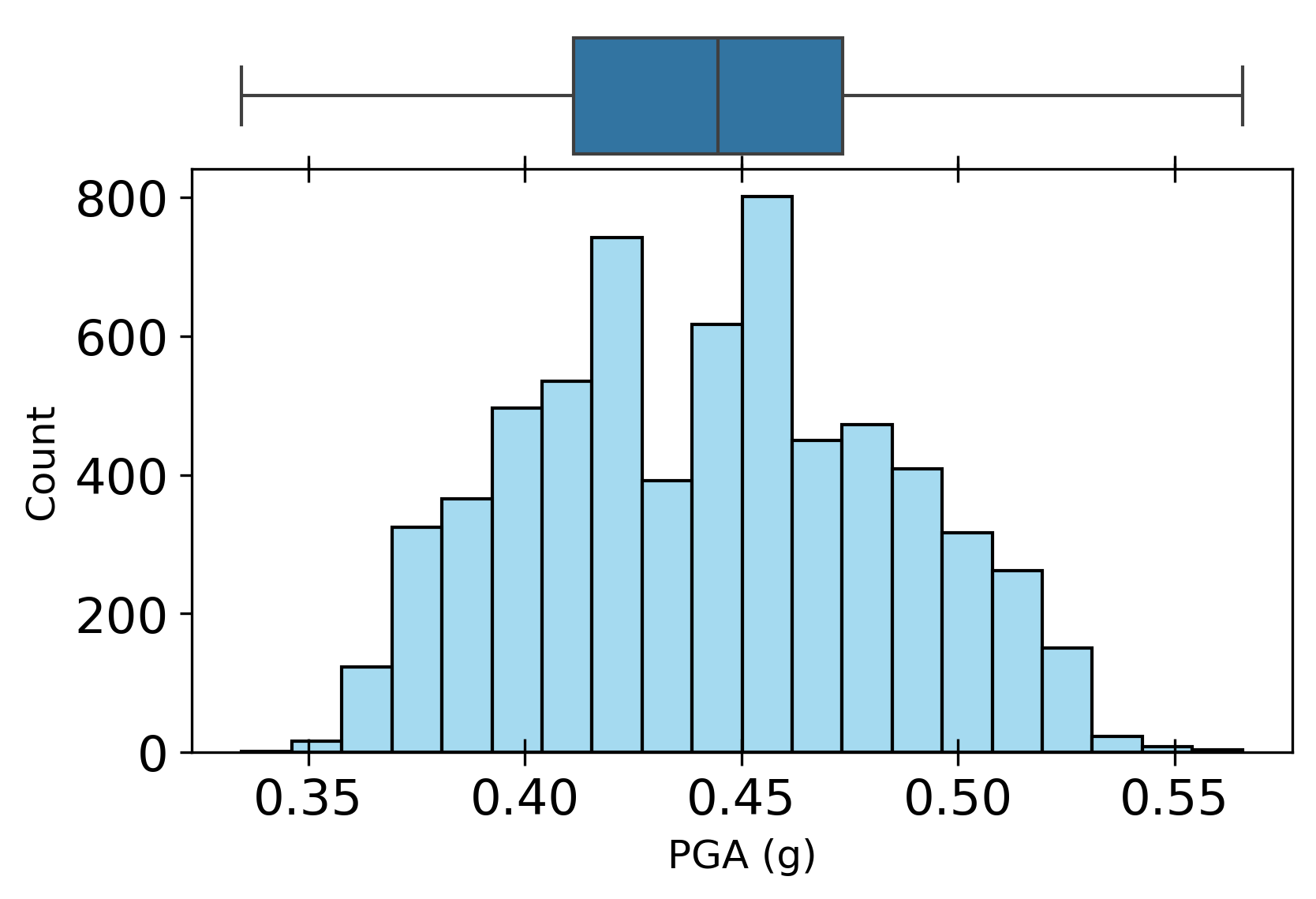}
        \caption{PGA}
        \label{fig:pga_distrib}
    \end{minipage}  
\setcounter{figure}{2}
\setcounter{subfigure}{2}
    \centering
    \begin{minipage}[b]{0.3\textwidth}
        \includegraphics[width=\linewidth]{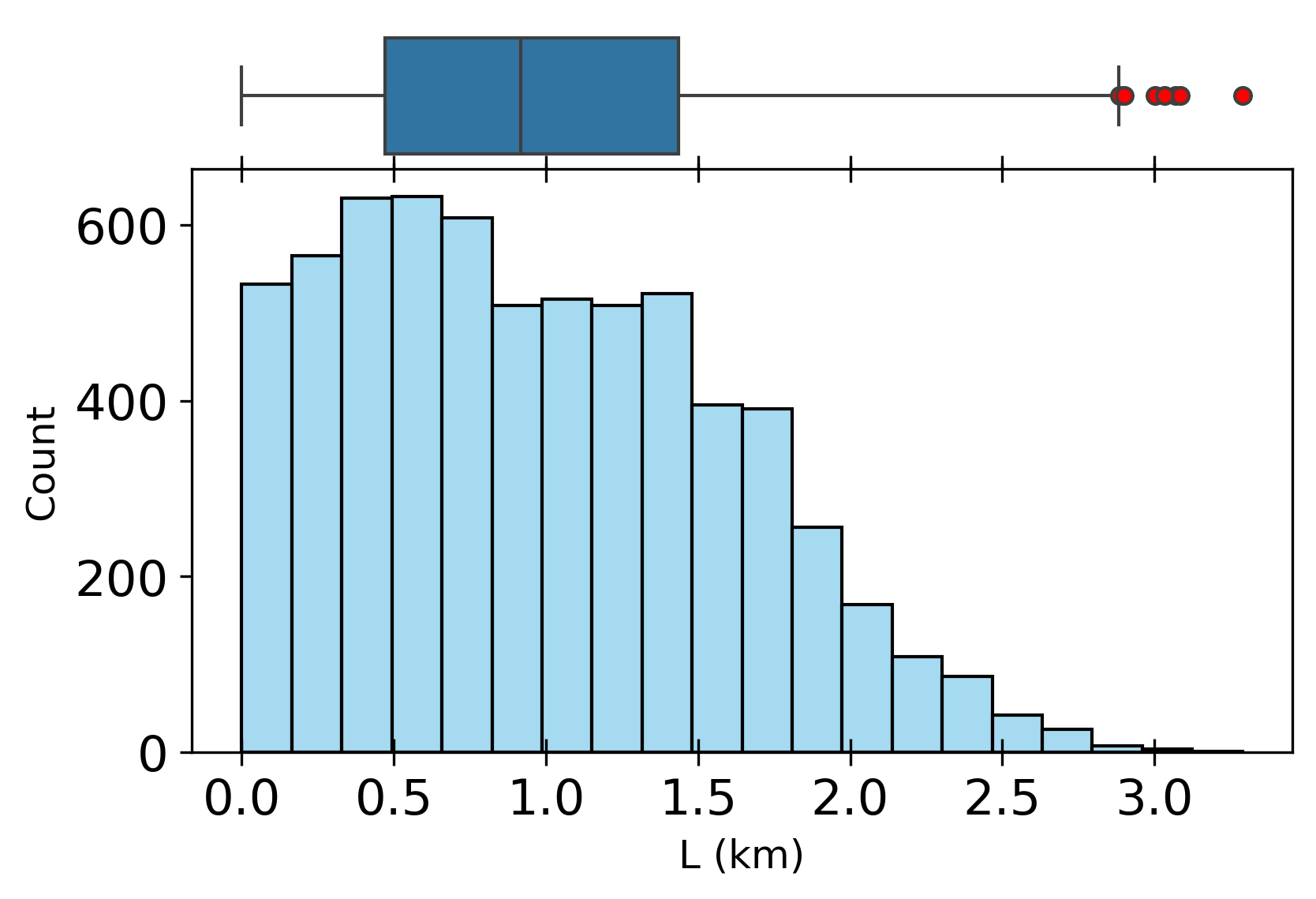}
        \caption{Distance to the river}
        \label{fig:l_distrib}
    \end{minipage} 
\setcounter{figure}{2}
\setcounter{subfigure}{3}
    \centering
    \begin{minipage}[b]{0.3\textwidth}
        \includegraphics[width=\linewidth]{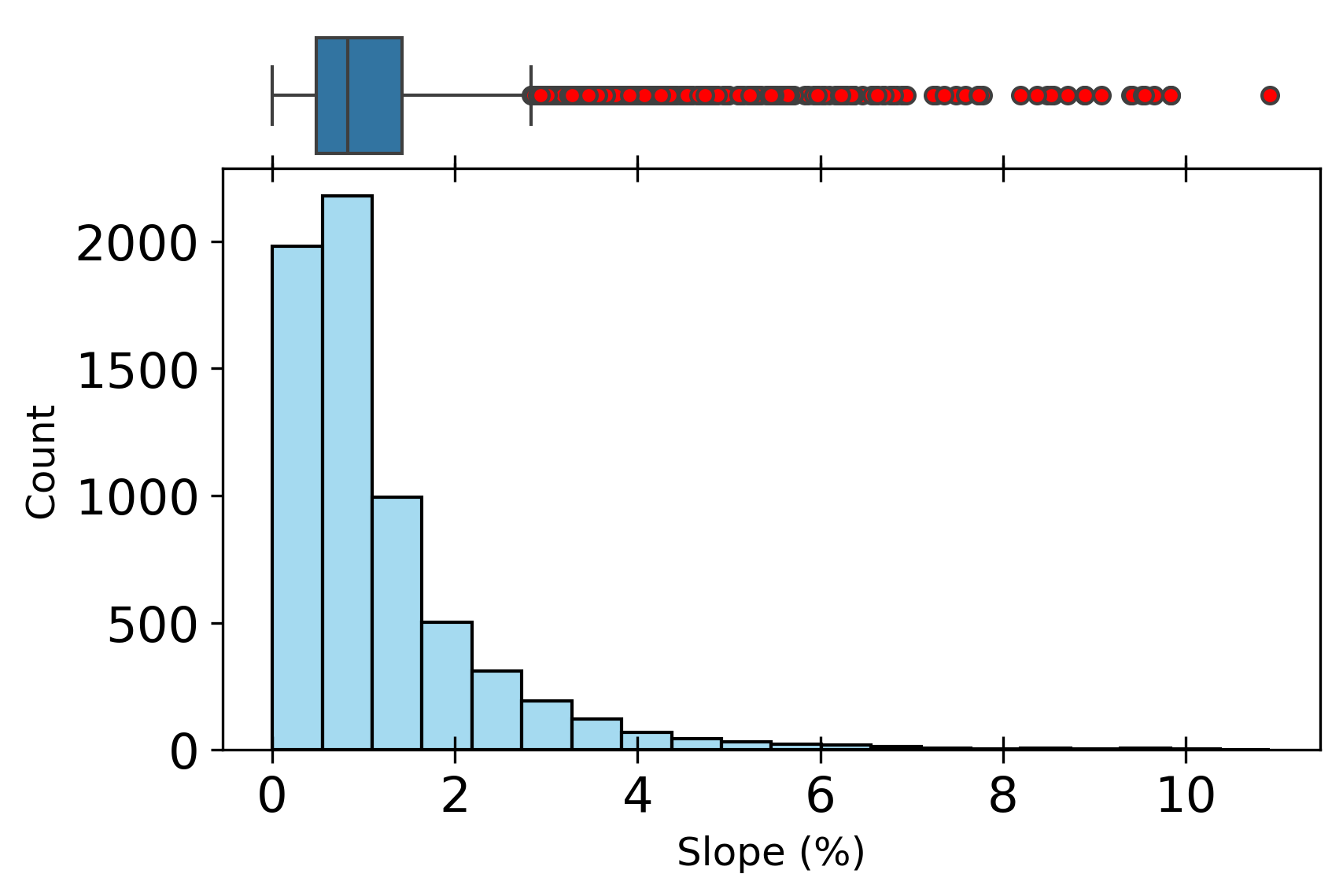}
        \caption{Slope}
        \label{fig:slope_distrib}
    \end{minipage} 
\setcounter{figure}{2}
\setcounter{subfigure}{4}
    \centering
    \begin{minipage}[b]{0.3\textwidth}
        \includegraphics[width=\linewidth]{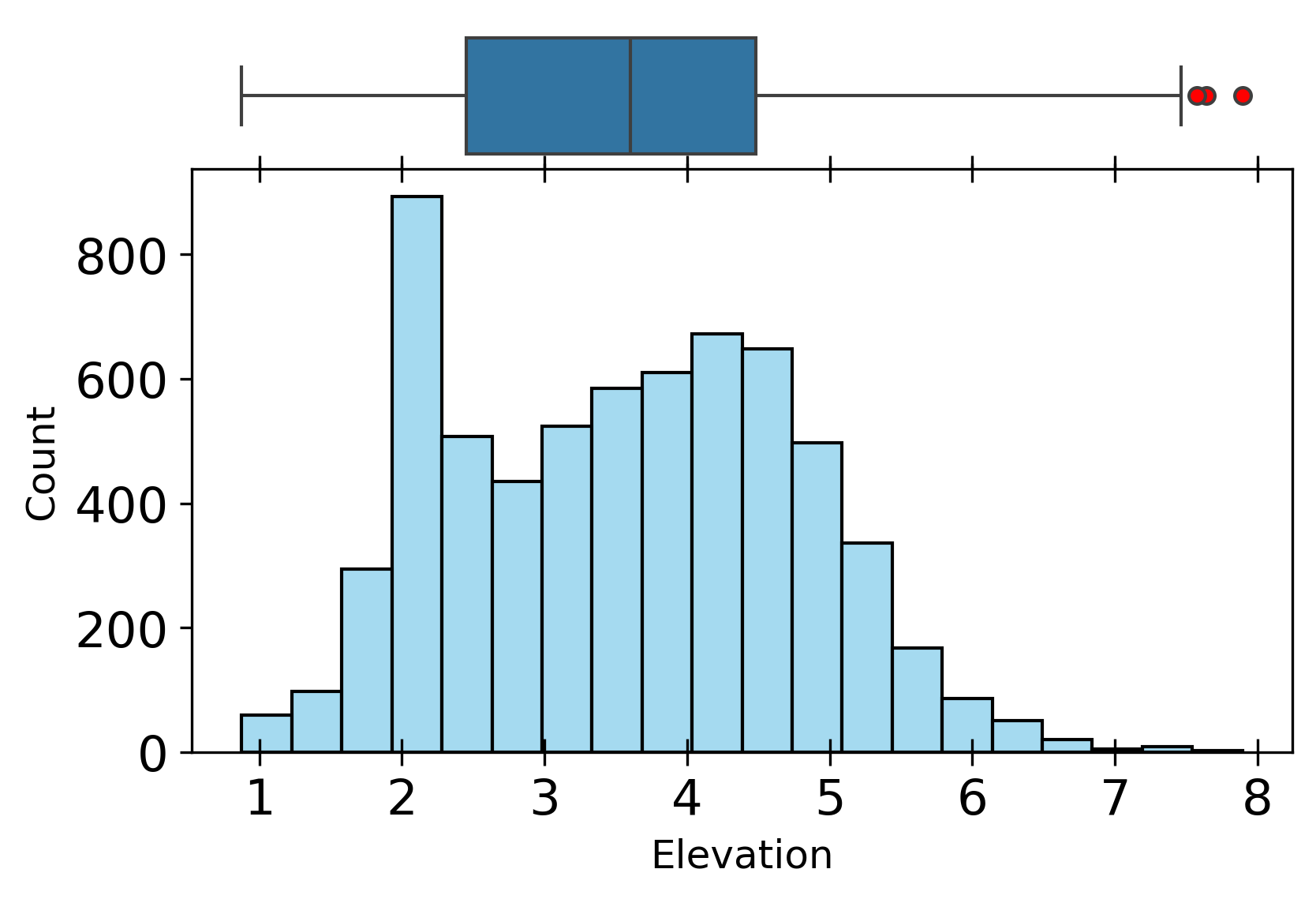}
        \caption{Elevation}
        \label{fig:el_distrib}
    \end{minipage} 
\setcounter{figure}{2}
\setcounter{subfigure}{-1}
    \caption{Distribution of input features and outliers analysis.
    \textbf{(A)} GWD
    \textbf{(B)} PGA
    \textbf{(C)} Distance to the river
    \textbf{(D)} Slope
    \textbf{(E)} Elevation
    }
    \label{fig:data_distrib}
\end{subfigure}

\section{Methodology}
\subsection{eXtreme Gradient Boosting}
\begin{figure}
\begin{center}
\includegraphics[width=10cm]{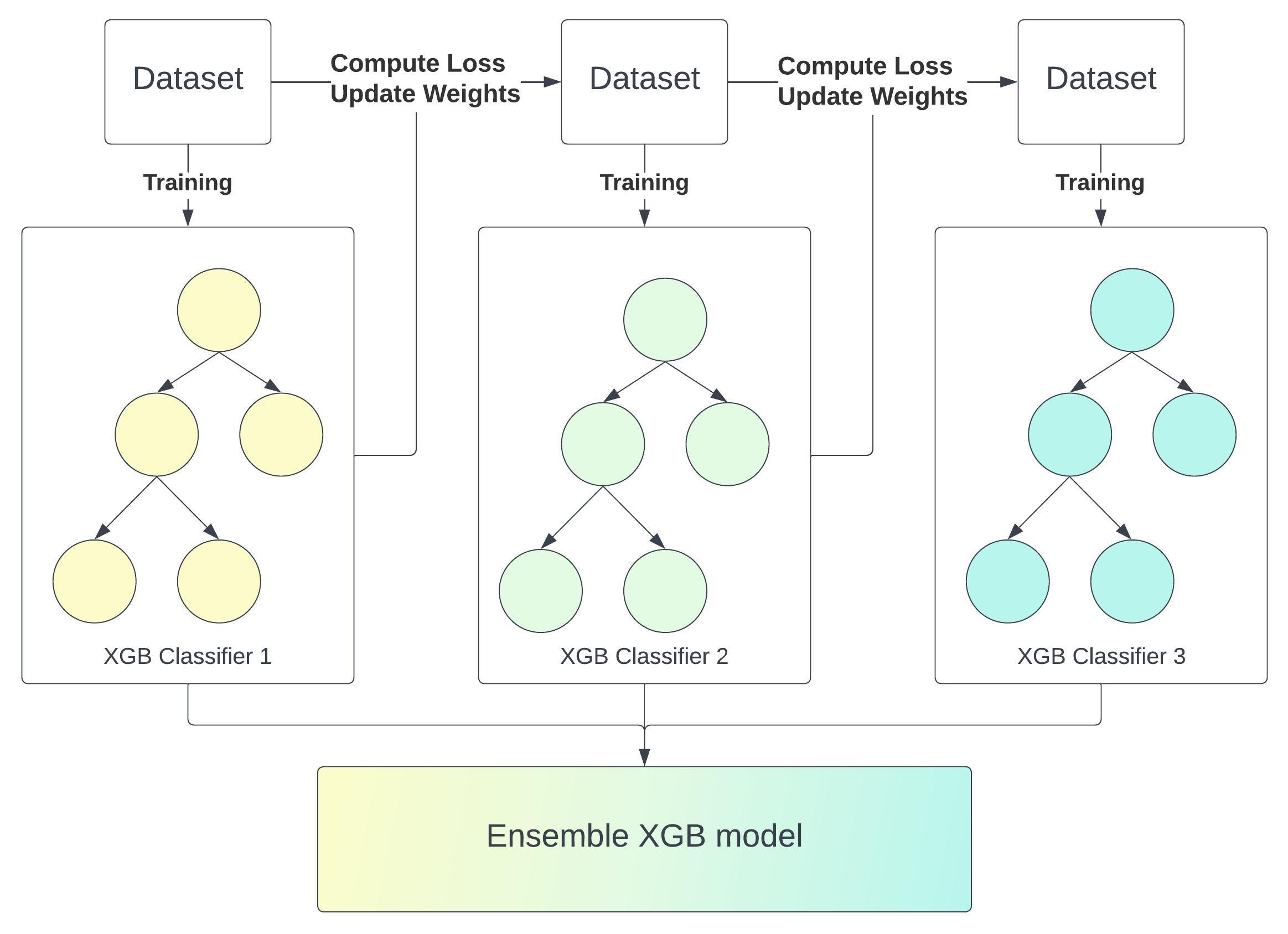}
\end{center}
\caption{Developing sequence of an XGB model}\label{fig:xgbtraining}
\end{figure}

XGB~\citep{Chen_2016} is an ensemble learning method extensively employed in various predictive modeling tasks. Fundamentally, XGB combines multiple weak classifiers, i.e., single decision trees, to construct a robust predictive model. This approach mirrors other tree-based ensemble methods like random forest, where the prediction is derived from the summation score of leaves across all trees. However, XGB distinguishes itself from the random forest in its tree generation process. While random forest concurrently builds all decision trees in the ensemble, XGB adopts a sequential approach through the gradient boosting technique (\Cref{fig:xgbtraining}). This methodology entails constructing new trees based on the loss (see~\Cref{eq:xgbloss}) computed from the previous trees. Additionally, XGB incorporates regularization terms into its loss function to mitigate overfitting, enhancing its predictive accuracy and generalization capabilities. 
\begin{equation}
Loss=\sum_{i=1}^{n}l(y_i+ y_i')+\lambda_1\sum_{j=1}^{t}\mid w_j\mid+ \lambda_2\sum_{j=1}^{t}w_j^2\label{eq:xgbloss}\,,
\end{equation}
where $l$ denotes the predictive loss function,  $y_i$ and $y_i'$ denote the target value and the predicted value of the training sample $i$, $w_j$ denotes the weights from tree $j$, and 1 and 2 denotes the coefficient for L1 and L2 regularization terms, respectively. Through this iterative refinement process, where the model continually improves based on the errors of previous iterations, XGB excels in achieving superior predictive performance compared to standalone weak classifiers or alternative ensemble methods. This capability enables XGB to capture intricate patterns and relationships inherent in the data, rendering it particularly suitable for tasks where predictive accuracy is paramount.

\subsection{SHAP}
In this study, we employ the SHapley Additive exPlanations (SHAP;~\citealp{lundberg-2017}) approach to provide insights into the decision-making process of the XGB model for predicting lateral spreading. 
The Shapley value~\citep{shapley-1953}, rooted in cooperative game theory, distributes the ``contributions" of players within a game equitably. It achieves this by calculating the marginal contribution when a player $i$ joins a player subset $S\subseteq F$, where $F$ represents the set of all players.~\Cref{eq:shap_value} illustrates the computation of the Shapley value $\phi_i$ for player $i$: 

\begin{equation}
\phi_i=\sum_{S\subseteq F}\frac{\mid S\mid!(\mid F\mid -\mid S\mid -1)!}{\mid F\mid!}[f(S\cup i)-f(S)]\label{eq:shap_value}\,,
\end{equation}
where $f$ denotes the model for evaluating the ``payout'' of a given subset of players and $[f(S\cup i)-f(S)]$ represents the marginal contribution when a player $i$  joins the player subset S. The coefficient $\mid S\mid !(\mid F\mid-\mid S\mid-1)!|F|!$, is applied to the marginal contribution, where $\mid F\mid$ denotes the total number of players, $\mid F\mid !$ represents the total permutations of all players, $\mid S\mid !$ accounts for the permutations of players already in the team before the player $i$ joins, and $(\mid F\mid-\mid S\mid-1)!$ represents the permutations of players who have not yet joined the team (excluding player $i$). This coefficient effectively weighs the impact of the permutation of a subset $S$. The Shapley value of player $i$ is then obtained as the weighted average of marginal contributions across all possible player subsets.

However, evaluating ``payout" from different subsets of input features requires retraining models, as the inputs vary across subsets. To address this challenge, SHAP offers an innovative approach by estimating Shapley values, termed SHAP values, through the conditional expectation of the original model denoted as $E[f(z)\mid z_S]$. Here, $z$ represents the model input with missing feature values for features not in the subset $S$, representing unknown information, while $z_S$ denotes the feature values for features within subset $S$. To compute the expected model output with missing features, we create sample instances with missing features filled by feature values sampled from the dataset. The random sampled feature values approximate the effect of missing. The expectation is then computed by averaging the model outputs of sampled instances. The SHAP value for each feature is calculated as the change in the expected model prediction when conditioning on that feature. 

\begin{figure}
\begin{center}
\includegraphics[width=10cm]{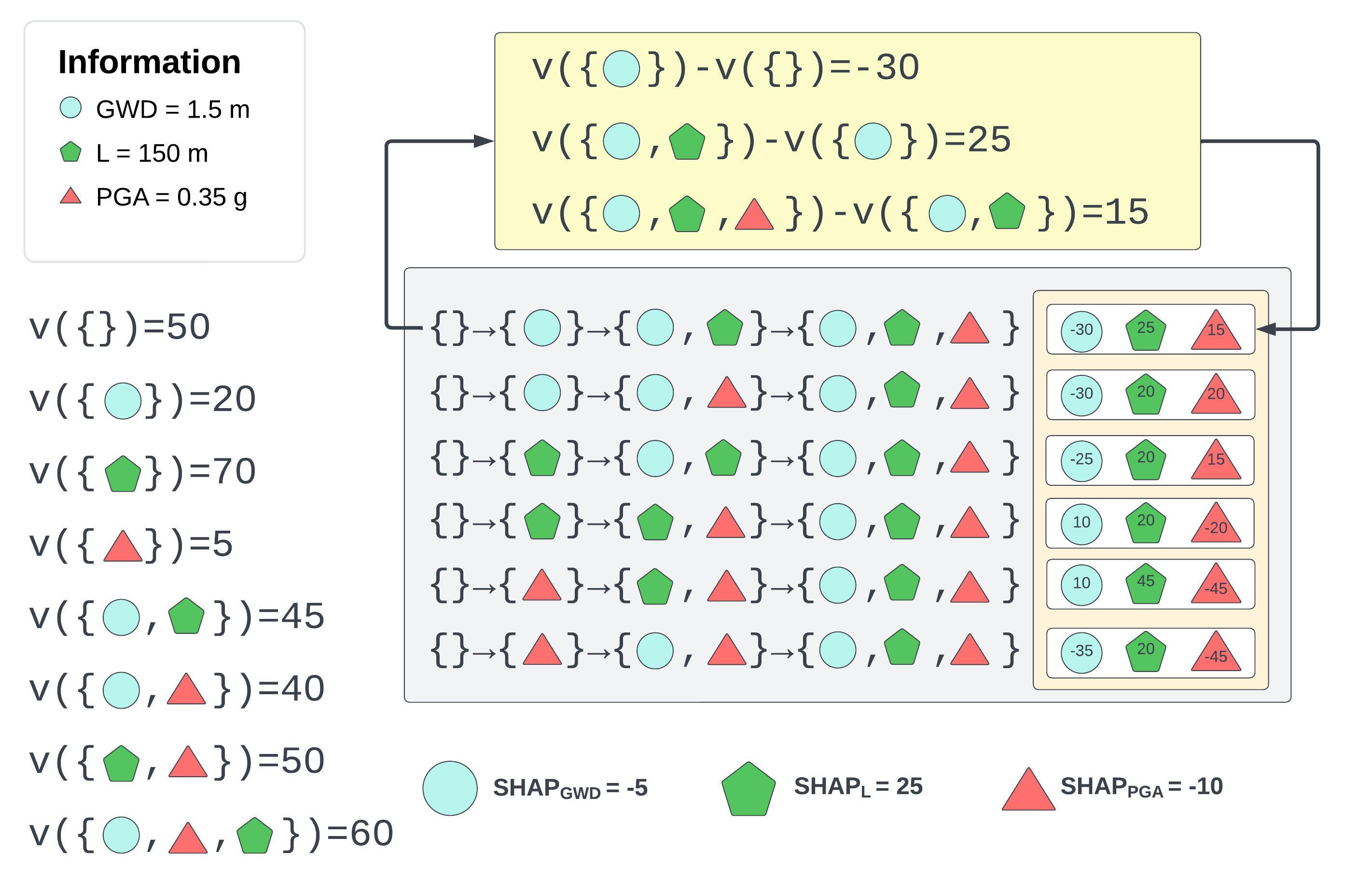}
\end{center}
\caption{An example of SHAP calculation for factors influencing lateral spreading.}
\label{fig:shap_example}
\end{figure}

In~\Cref{fig:shap_example}, we provide an example of SHAP calculation of predicting the risk of lateral spreading based on three factors: L (Distance to the river), PGA (Peak Ground Acceleration), and GWD (Groundwater Depth). Consider a model that predicts the risk of lateral spreading with a risk score ranging from 0 (no risk) to 100 (highest risk). The model assigns a risk score of 60 for a site with these characteristics: L = 150 m, PGA = 0.35g, and GWD = 1.5 meters.

The SHAP analysis begins with a baseline risk score of 50 ($v\{\} = 50$), here $v$ represents a function that returns the expected risk, which is the model’s output without specific data for the factors, based only on the training data. When assessing the individual impact of each factor, we find that introducing L (150 m) alone results in a risk score increase to 70, indicating a positive SHAP value for L ($SHAP_L:v\{L\}-v\{\} = 70 - 50 = 20$). However, when we introduce PGA (0.35g) alone, the risk score decreases to 5, suggesting a negative SHAP value for PGA ($SHAP_{PGA}:v\{PGA\}-v\{\}=5-50=-45$), implying that this PGA level reduces the model's perceived risk. Similarly, when GWD (1.5 m) is introduced alone, the risk score decreases to 20, indicating a negative SHAP value for GWD ($SHAP_{GWD}:v\{GWD\}-v\{\} = 20-50=-30$).

Next, we evaluate the pairwise combinations. With both L and GWD, the risk score increases to 45, showing a combined positive effect, even though GWD alone has a negative contribution. With L and PGA, the risk score reaches 50, demonstrating a lower negative contribution from PGA when considered alongside L, indicating an interaction effect where PGA's negative impact is mitigated. With PGA and GWD, the risk score is 40, suggesting that their combined effect is less negative than their individual contributions.

Considering all three factors, the model predicts a risk score of 60. The combined influence of L, PGA, and GWD results in a total increase of 10 points from the baseline, with the interplay between the factors affecting their individual contributions.

Finally, we average the contributions across all permutations to calculate the SHAP values. This process yields average SHAP values, with positive values indicating factors increasing the risk and negative values indicating factors that reduce the risk score. The SHAP values in order of highest to lowest importance are $SHAP_L = 25$ (strongest positive impact), $SHAP_{PGA}= -10$ (negative impact, but less when combined with L), and $SHAP_{GWD} = -5$ (negative impact).

These SHAP values explain the model’s risk prediction for lateral spreading by illustrating how each factor contributes and how their interactions affect the overall risk score. The presence of negative SHAP values helps to understand the conditions under which certain factors may decrease the perceived risk, providing a more nuanced explanation of the model's behavior.

One of the key advantages of SHAP is its model-agnostic nature. Regardless of the model's underlying structure or type, whether a tree-based, neural network, or any other functional model, SHAP can compute Shapley values by analyzing the input permutations and model outputs. Another notable advantage of SHAP is its ability to provide global and local explanations. While other explainers, such as permutation feature importance ~\citep{breiman-2001} and Partial Dependence Plots (PDP;~\citealp{friedman-2001}) offer only global explanations that describe the average behavior of the model. SHAP and methods like LIME can provide explanations for individual predictions, thereby offering local insights. Furthermore, SHAP explanations exhibit a high degree of stability. The results remain consistent when rerunning the SHAP explainer due to its computation of Shapley values. In contrast, LIME explanations may vary as they involve creating a surrogate model through a sampling process, which can lead to inconsistencies.

However, one potential drawback of SHAP is its computational time. The intensive computations required for a single instance, as demonstrated in the example referred to in~\Cref{fig:shap_example}, can make it impractical to utilize SHAP for a large number of instances or instances with numerous features. Nevertheless, given the manageable dataset size in our study, with only 6500 instances and five to nine features, the computation time remains within a few minutes, allowing us to select SHAP as the explainer for our analysis confidently.

\section{Model Development}
We develop an XGBoost classifier, Model A, trained on five features (PGA, GWD, L, slope, and elevation)  to predict lateral spreading. The dataset is split into three subsets with a ratio of 80-5-15 for training, validation, and testing.~\Cref{tab:datasplit} presents the distribution of Yes/No data across each dataset. In the training data, the ratio of Yes to No instances is 42-58, while in the validation data, it is 46-54, and in the testing data, it is 43-57. Although there is slight variation in the proportion of Yes and No instances across datasets due to randomness, it closely mirrors the ratio observed in the total dataset (i.e., 42-58), with no significant imbalance detected.

Model A is trained on training data using the log loss function (\Cref{eq:logloss}) as loss function. The equation for the log loss function is represented as:

\begin{equation}
l=-\frac{1}{n}\sum_{i=1}^{n}y_i\cdot log(y_i')+(1-y_i)\cdot log(1-y_i')\,,\label{eq:logloss}
\end{equation}
where $n$ denotes the number of training data points,  $y_i$ denotes the binary label indicating lateral spreading for data point $i$ (0 for no lateral spreading, 1 for lateral spreading), $y_i'$ represents the predicted probability, ranging from 0 to 1, generated by Model A.

We optimize the configuration of Model A using cross-validation preprocessing.~\Cref{tab:model_info} shows the optimized parameters for Model A. The training also restricts the depth of the decision trees in the ensemble to avoid over-fitting the data because our interest is in explaining the predictions rather than the most accurate model possible. 
\begin{table}
    \centering
    \caption{Number of Yes/No data in each dataset.}
    \begin{tabular}{ccccc}
        \toprule
        Dataset & Yes lateral spread & \multicolumn{1}{l}{No lateral spread} \\
        \midrule
        Total & 2752 (42.31\%) & 3753 (57.69\%) \\
        Train & 2182 (41.93\%) & 3022 (58.07\%) \\
        Val   & 150 (46.01\%)  & 176 (53.99\%)  \\
        Test  & 420 (43.08\%)  & 555 (56.92\%)  \\
        \bottomrule
    \end{tabular}
    \label{tab:datasplit}
\end{table}
\begin{table}
    \centering
    \caption{Model information: Training set features, model configurations, and performance.}
    \begin{tabular}{ccccc}
        \toprule
        & & Model A & Model B & Model C\\
        \midrule
        &Distance from the river (L) in km & \checkmark & \checkmark & \checkmark\\
        &Peak Ground Acceleration (PGA) in g & \checkmark & \checkmark & \checkmark\\
        &Elevation in m & \checkmark & \checkmark & \checkmark\\
        &Groundwater Depth (GWD) in m & \checkmark & \checkmark & \checkmark\\
        Input Feature&Slope angle in \% & \checkmark & \checkmark & $\bigcirc$\\
        &$I_{c}$ Median & $\bigcirc$ & \checkmark & \checkmark\\
        &$q_{c1Ncs}$ Median & $\bigcirc$ & \checkmark & $\bigcirc$\\
        &$I_c$ STD & $\bigcirc$ & \checkmark & \checkmark\\
        &$q_{c1Ncs}$ STD & $\bigcirc$ & \checkmark & $\bigcirc$\\ 
        \midrule
        &Loss function & \multicolumn{3}{c}{log-loss}\\
        &Max Tree depth & \multicolumn{3}{c}{6}\\
        &$\lambda_1$ & \multicolumn{3}{c}{0.0}\\
        Parameter&$\lambda_2$ & \multicolumn{3}{c}{1.0}\\
        &Early stop & \multicolumn{3}{c}{50}\\
        &Learning rate & \multicolumn{3}{c}{0.300}\\
        &Number of estimators & 52 & 29 & 90\\
        \midrule
        &Training (\%) & 94.37 & 91.99 & 98.33\\
        Accuracy & Validation (\%) & 87.12 & 86.50 & 88.04\\
        &Testing (\%) & 84.21 & 82.77 & 85.54\\
        \bottomrule
    \end{tabular}
    \label{tab:model_info}
\end{table}

\section{Results}

Model A yields 87.12\% accuracy on the validation data and 84.21\% on the testing data (see~\Cref{tab:model_info}).~\Cref{fig:model_a_cm} shows the confusion matrix on the testing data, showing the number of cases in four categories: true positive (TP), true negative (TN), false positive (FP), and false negative (FN). The model correctly classifies 488 no-lateral-spreading cases (50.1\% TN) and 333 lateral-spreading cases (34.2\% TP). The model misclassified 67 no-lateral-spreading cases (6.9\%) as lateral spreading cases (FP) and 87 lateral-spreading cases (8.9\%) as no-lateral-spreading cases (FN).~\Cref{fig:model_a_predictionmap} shows the spatial distribution of these four categories with different markers. We highlight two regions of false prediction to discuss in the result section: Region A for FNs and Region B for FPs.  In Region A, only 25\% of lateral spreading cases are correctly predicted in the testing data, while the predictions in the training data match 50\% of the cases.  Meanwhile, Model A only predicts 40\% of no-lateral-spreading cases in the Region B testing dataset. This underscores the importance of unpacking the 'black box' and the model training dataset to understand the underlying physics or mechanisms learned by the model.

\Cref{fig:model_a_prob} shows the distribution of predictive probabilities within the testing dataset. Predictive probability serves as an indicator of the classification predicted by the model, where a value exceeding 0.5 signifies a prediction of lateral spreading. In contrast, a value below 0.5 indicates a classification of no-lateral spreading. Additionally, the predictive probability is the model's confidence in its predictions. A probability closing 1.0 or 0.0 suggests high confidence, whereas a probability near 0.5 indicates that the model is uncertain.

In~\Cref{fig:probs_tntp}, the distribution of TPs is depicted in dark gray, peaking at a probability of 0.99 with 63 instances (19\% of TPs). 51.9\% of TP cases fall within the range of 0.9 to 1.0, indicating high confidence in the TP category. A similar pattern of high confidence is observed in the distribution of TNs. In contrast, the predictive probabilities associated with FPs and FNs are closer to 0.5, as shown in~\Cref{fig:probs_fnfp}. Notably, 34\% of FPs fall within the range of 0.5 to 0.6, while only 10.8\% of TPs are within the same range. No FPs exhibit high predictive probabilities exceeding 0.96. Similarly, 19.5\% of FNs fall within 0.4 to 0.5, in contrast to only 5\% of TNs. This comparison underscores that the model tends to be more confident in its correct predictions, as evidenced by higher predictive probabilities while displaying less confidence in incorrect predictions, as indicated by the clustering around 0.5.
\begin{subfigure}
\setcounter{figure}{5}
\setcounter{subfigure}{0}
    \centering
    \begin{minipage}[b]{0.35\textwidth}
        \includegraphics[width=\linewidth]{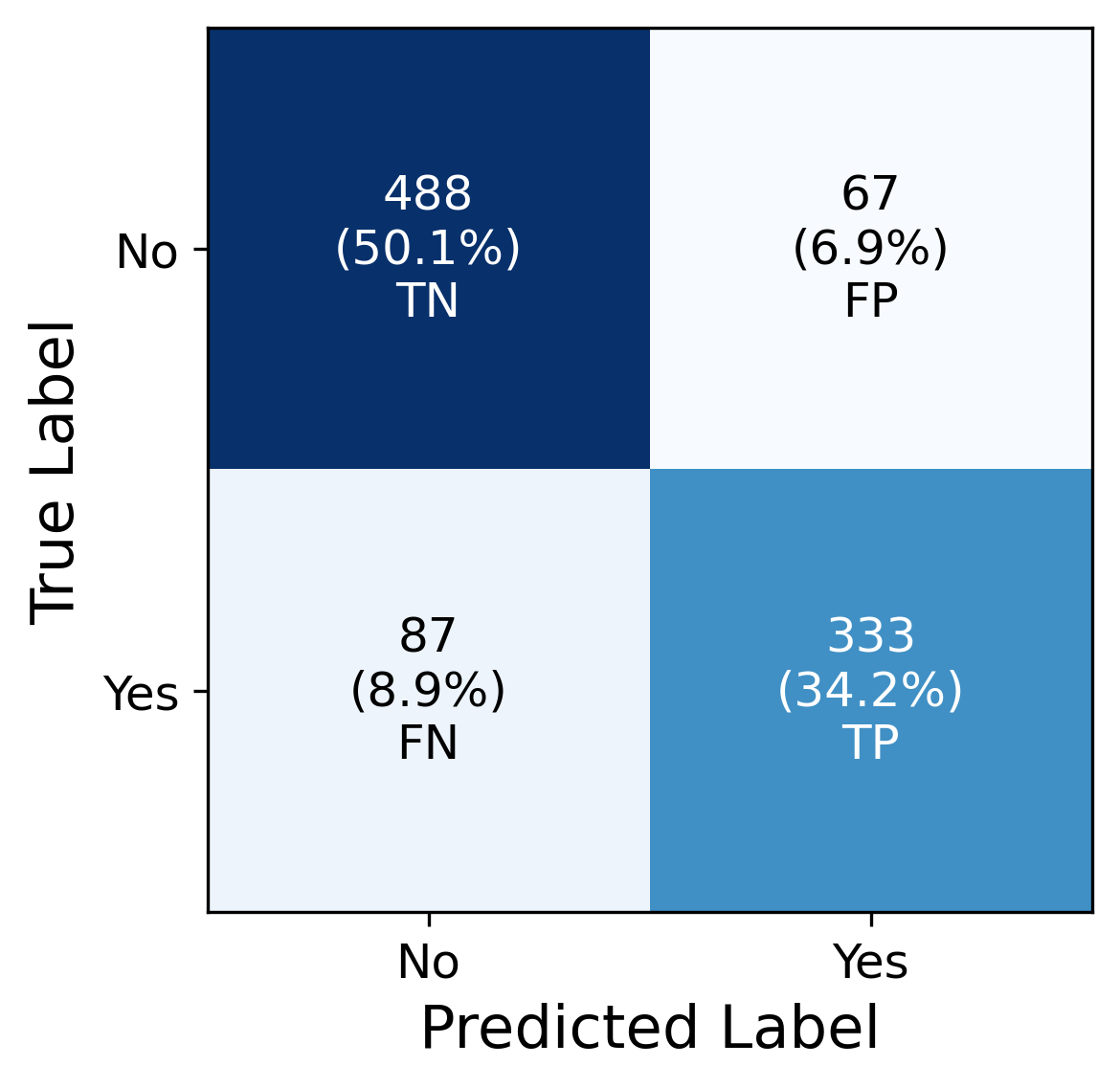}
        \caption{Confusion matrix of testing data}
        \label{fig:model_a_cm}
    \end{minipage}  
\setcounter{figure}{5}
\setcounter{subfigure}{1}
    \begin{minipage}[b]{0.5\textwidth}
        \includegraphics[width=\linewidth]{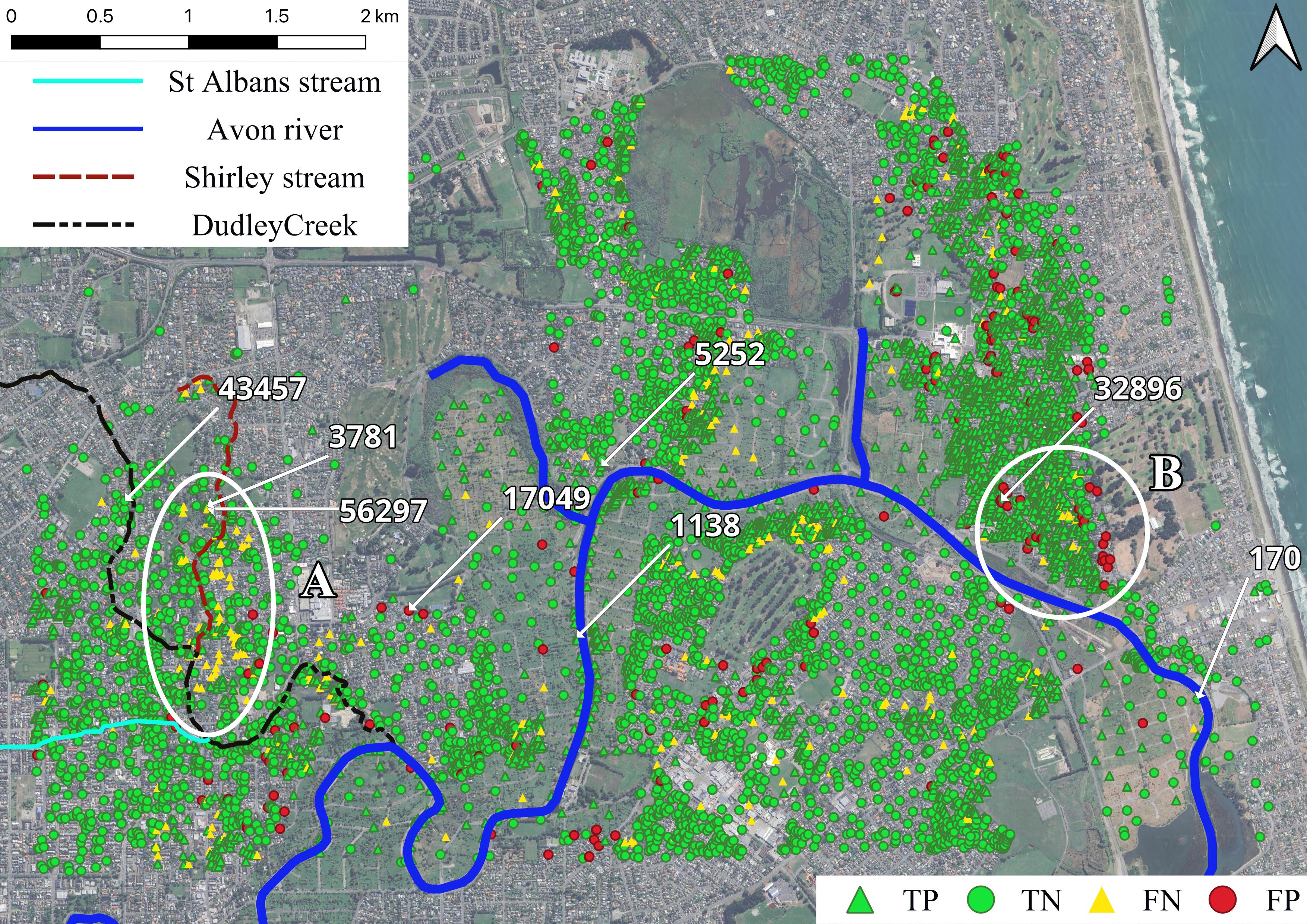}
        \caption{Spatial distribution of predictions}
        \label{fig:model_a_predictionmap}
    \end{minipage}
\setcounter{figure}{5}
\setcounter{subfigure}{-1}
    \caption{Performance of the Model A:
    \textbf{(a)} Confusion matrix of testing data.
    \textbf{(b)} Spatial distribution of predictions of the whole dataset.
    }
    \label{fig:model_a_performance}
\end{subfigure}
\begin{subfigure}
\setcounter{figure}{6}
\setcounter{subfigure}{0}
    \centering
    \begin{minipage}[b]{0.4\textwidth}
        \includegraphics[width=\linewidth]{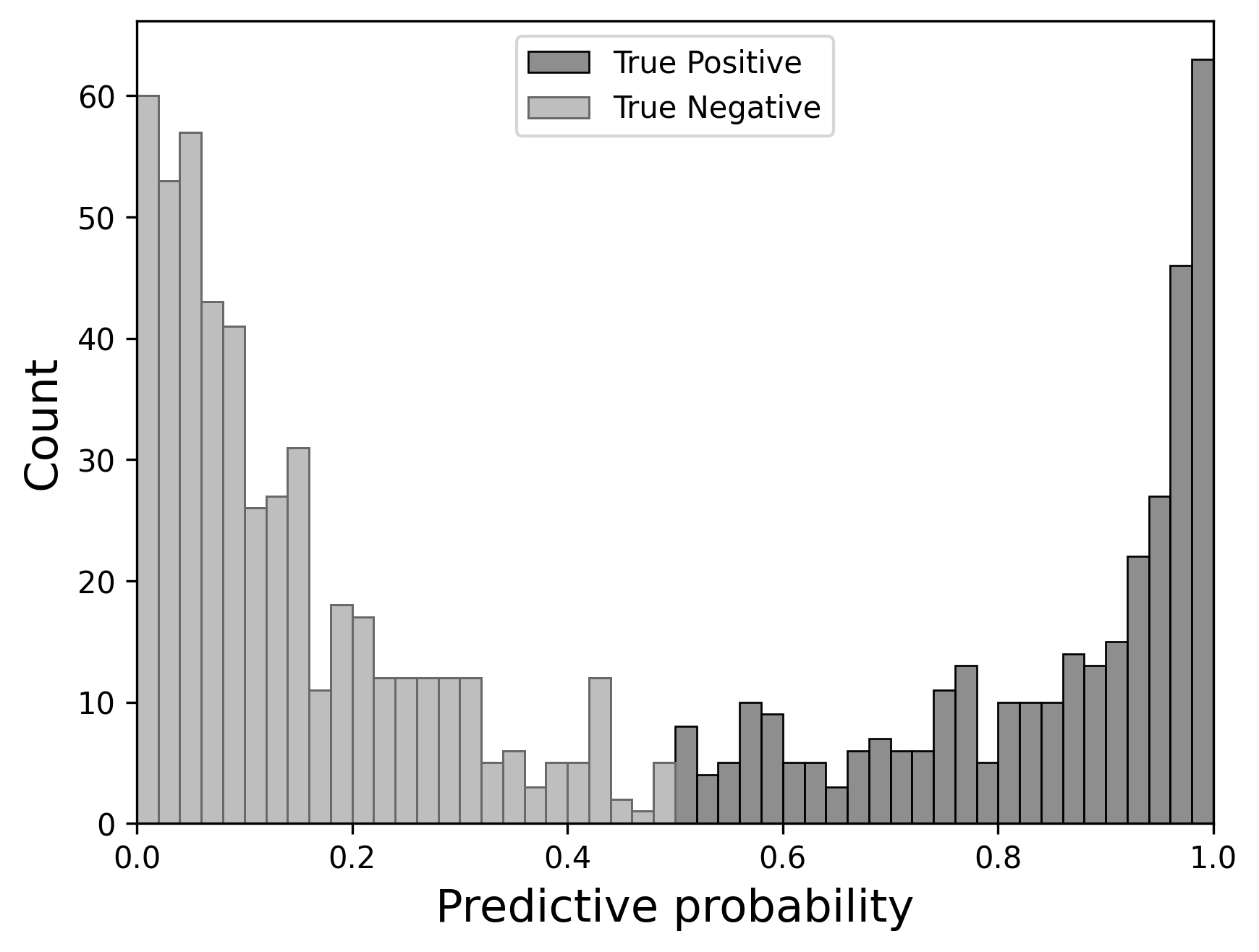}
        \caption{TP and TN}
        \label{fig:probs_tntp}
    \end{minipage}  
\setcounter{figure}{6}
\setcounter{subfigure}{1}
    \begin{minipage}[b]{0.4\textwidth}
        \includegraphics[width=\linewidth]{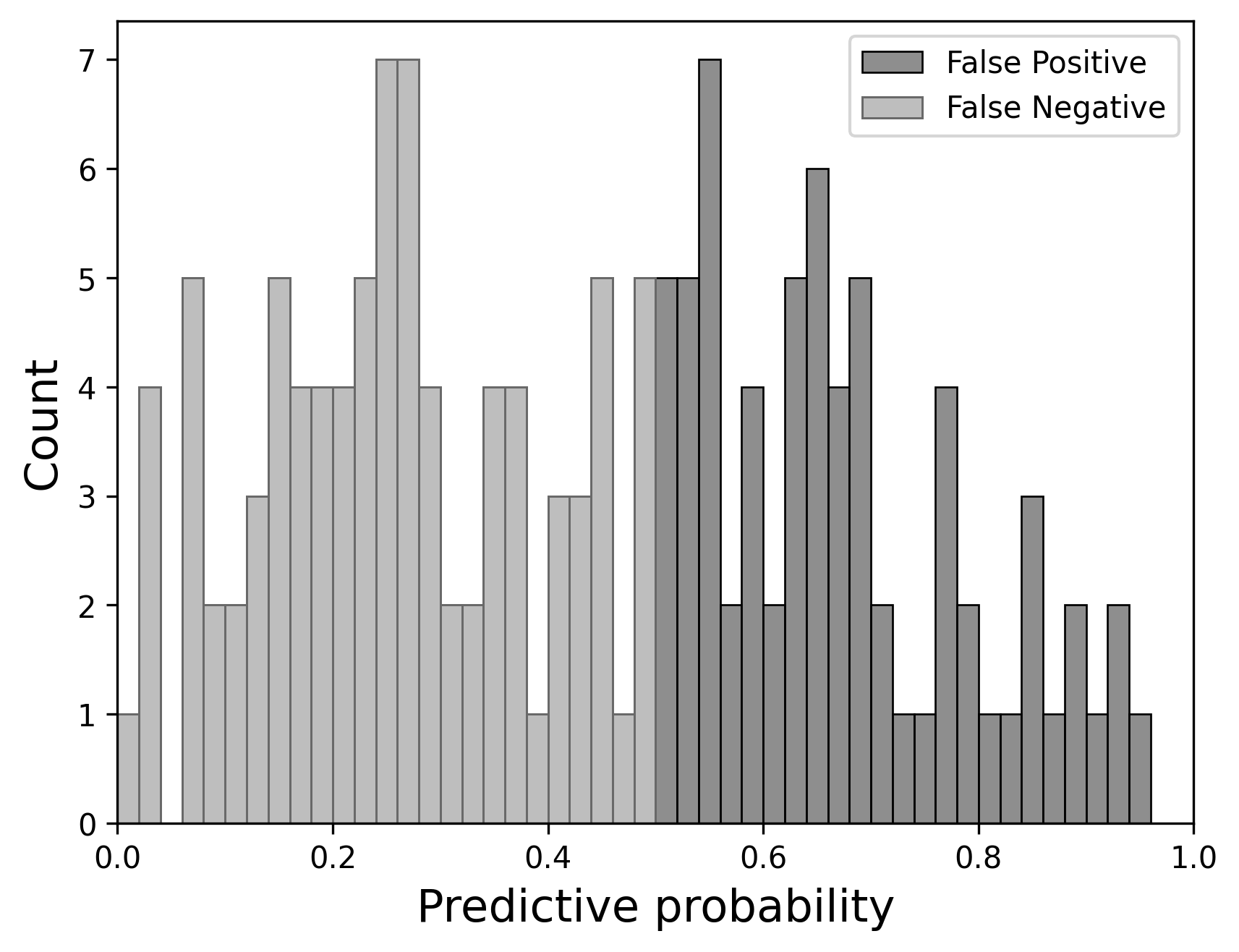}
        \caption{FP and FN}
        \label{fig:probs_fnfp}
    \end{minipage}
\setcounter{figure}{6}
\setcounter{subfigure}{-1}
    \caption{The predictive probability distribution of Model A in the testing set is categorized as
    \textbf{(a)} TP and TN
    \textbf{(b)} FP and FN.
    }
    \label{fig:model_a_prob}
\end{subfigure}

\subsection{Local explanations}
\subsubsection{True positives and true negatives}
We focus on specific sites to demonstrate the predictive ability of XGBoost Model A and explain the predictions. Consider Site 1138 (see~\Cref{fig:model_a_predictionmap}), which is situated near the river (L = 6 m) and is thus prone to lateral spreading. Model A correctly predicts lateral spreading with a 93.1\% probability.~\Cref{fig:model_a_tp} illustrates the SHAP values as a waterfall plot for various features influencing this prediction. A positive SHAP value suggests a higher likelihood of lateral spreading, whereas a negative value indicates no lateral spreading. Site 1138 has a high positive SHAP value of 2.6.03, demonstrating a significant probability of lateral spreading. SHAP prioritizes the five input features based on their predictive importance for this site, with proximity to the river (L = 6 m) being the most crucial and PGA (0.444 g) the least. The SHAP explanations suggest that the site's location near the river and a low GWD  of 1.055 m are the reasons the site will likely experience lateral spreading. This finding aligns with established domain knowledge.

On the other hand, Site 43457 (see~\Cref{fig:model_a_predictionmap}), situated further from the river, represents a different scenario. The SHAP waterfall plot (see~\Cref{fig:model_a_tn}) indicates that its location (L = 2619 m from the river) is a critical factor in predicting the absence of lateral spreading. The prediction aligns with our field observations in the training data, where fewer than 17.5\% of sites located more than 2 km from the river experienced liquefaction. The site's high elevation, low GWD, and low slope (having a minor positive influence) contribute to this prediction, resulting in an overall negative SHAP value of -3.713. Note that the training dataset is slightly unbalanced relative to the yes/no observations of lateral spreading, with the ratio of no sites to yes sites, termed the scale positive weight, equal to 1.38. The imbalance in the dataset leads to an initial bias in the expected SHAP value ($E[f(x)]=v\{\} = -0.369$); before considering any features, the model expects a given site not to experience lateral spreading, i.e., a baseline prediction is no lateral spreading.
\begin{subfigure}
\setcounter{figure}{7}
\setcounter{subfigure}{0}
    \centering
    \begin{minipage}[b]{0.45\textwidth}
        \includegraphics[width=\linewidth]{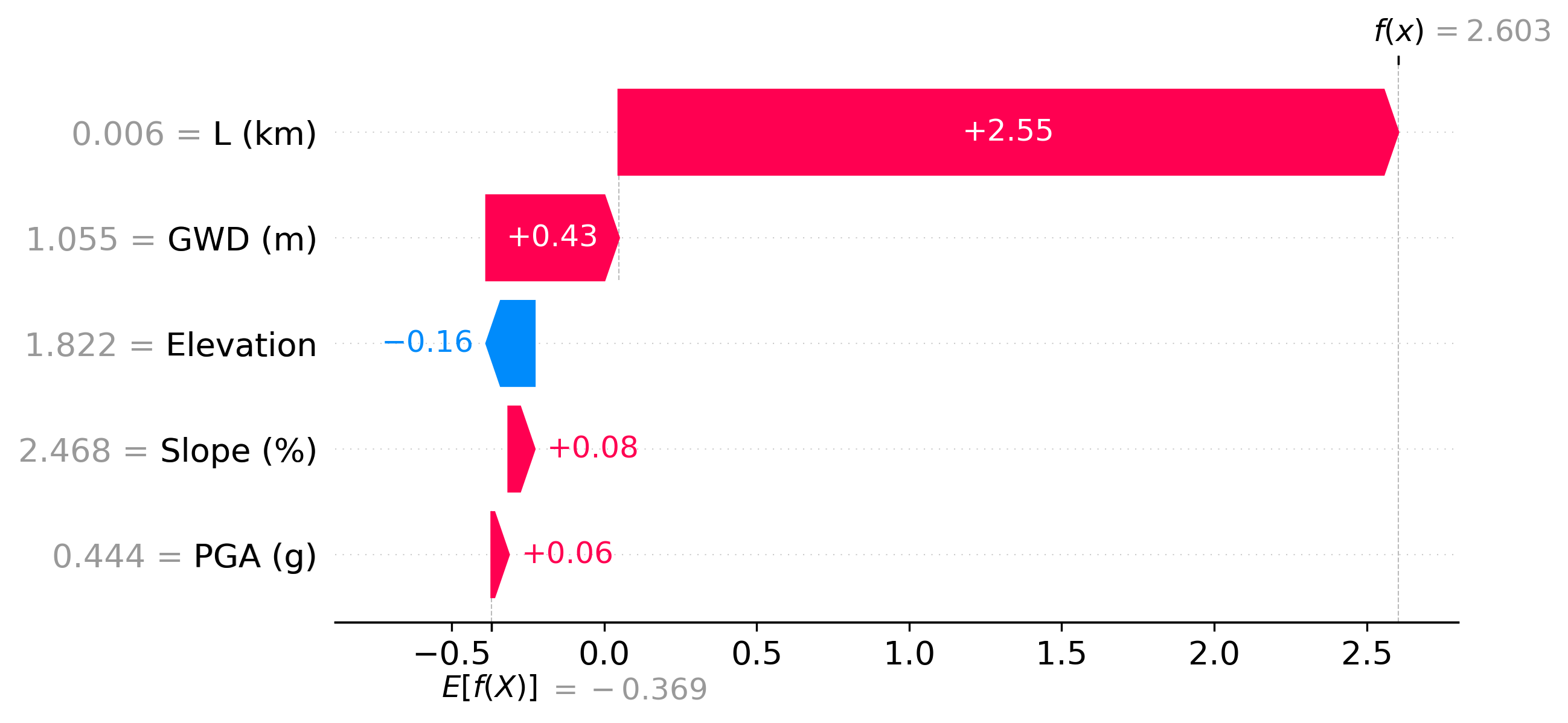}
        \caption{Site 1138 (True positive); 0.941 m displacement; 93.10\%}
        \label{fig:model_a_tp}
    \end{minipage}  
\setcounter{figure}{7}
\setcounter{subfigure}{1}
    \begin{minipage}[b]{0.45\textwidth}
        \includegraphics[width=\linewidth]{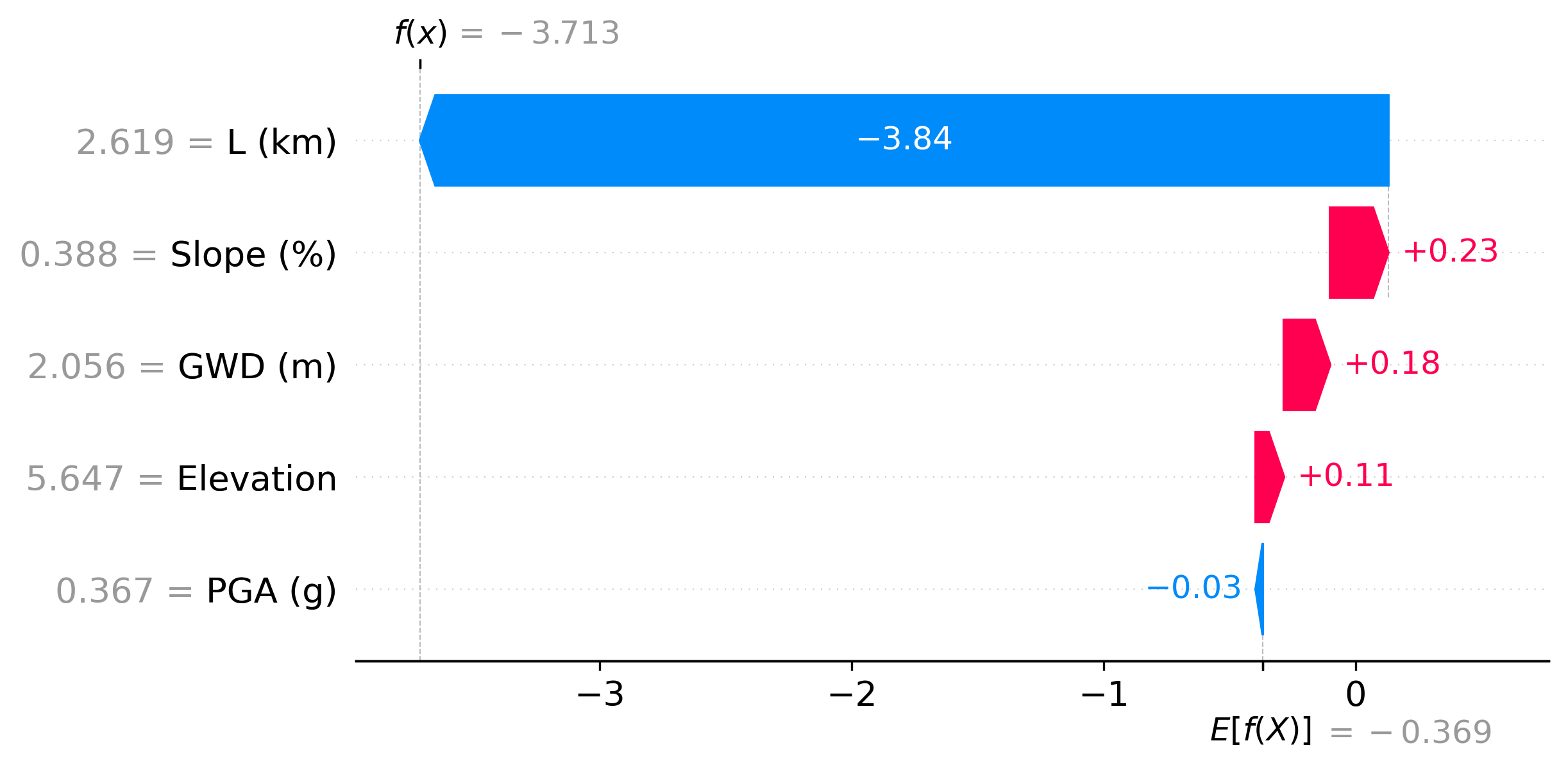}
        \caption{Site 43457 (True negative); 0.214 m displacement; 2.38\%}
        \label{fig:model_a_tn}
    \end{minipage}
\setcounter{figure}{7}
\setcounter{subfigure}{2}
    \begin{minipage}[b]{0.45\textwidth}
        \includegraphics[width=\linewidth]{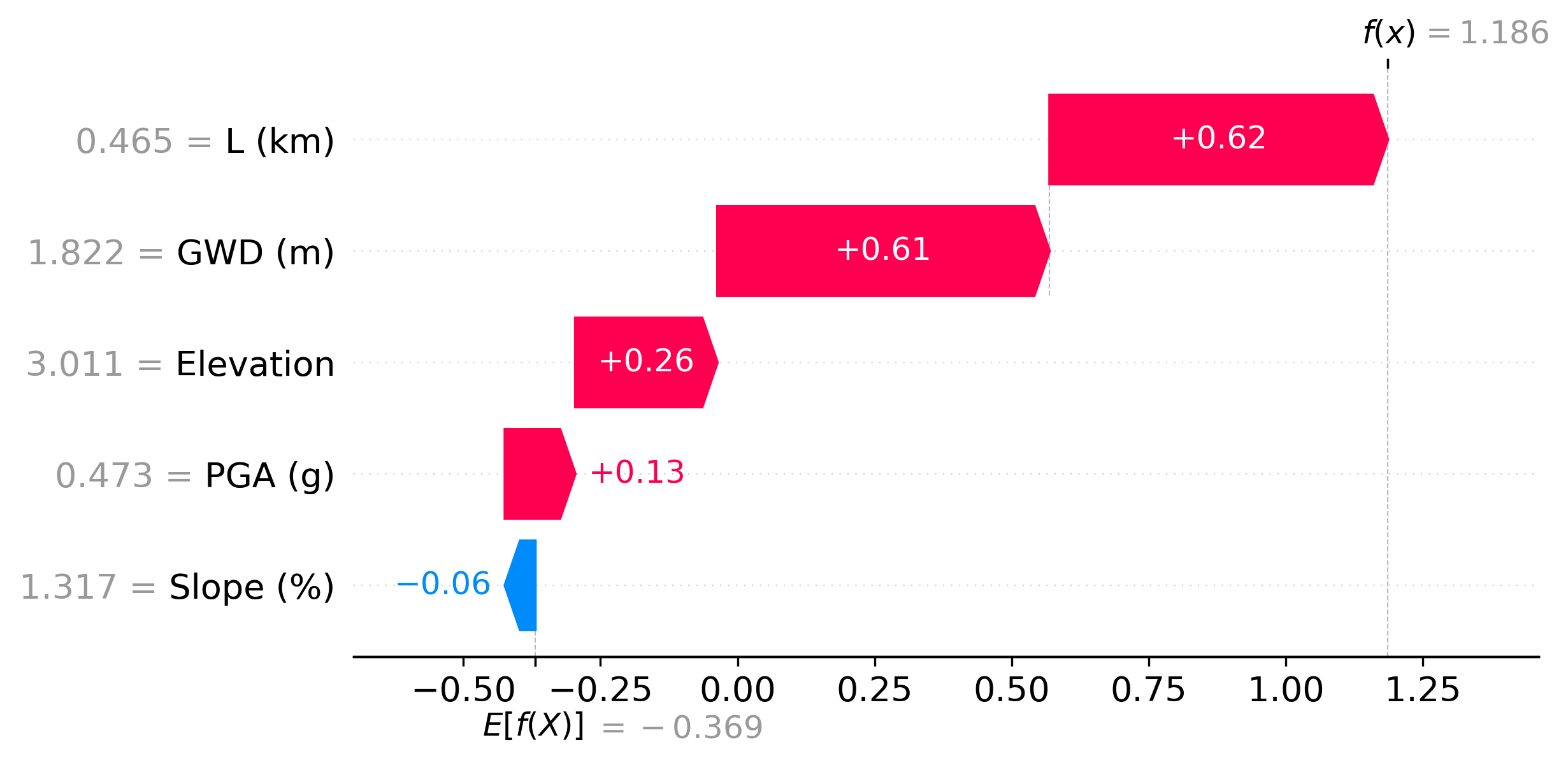}
        \caption{Site 32896 (False positive); 0.259 m displacement; 76.60\%}
        \label{fig:model_a_fp}
    \end{minipage}
\setcounter{figure}{7}
\setcounter{subfigure}{3}
    \begin{minipage}[b]{0.45\textwidth}
        \includegraphics[width=\linewidth]{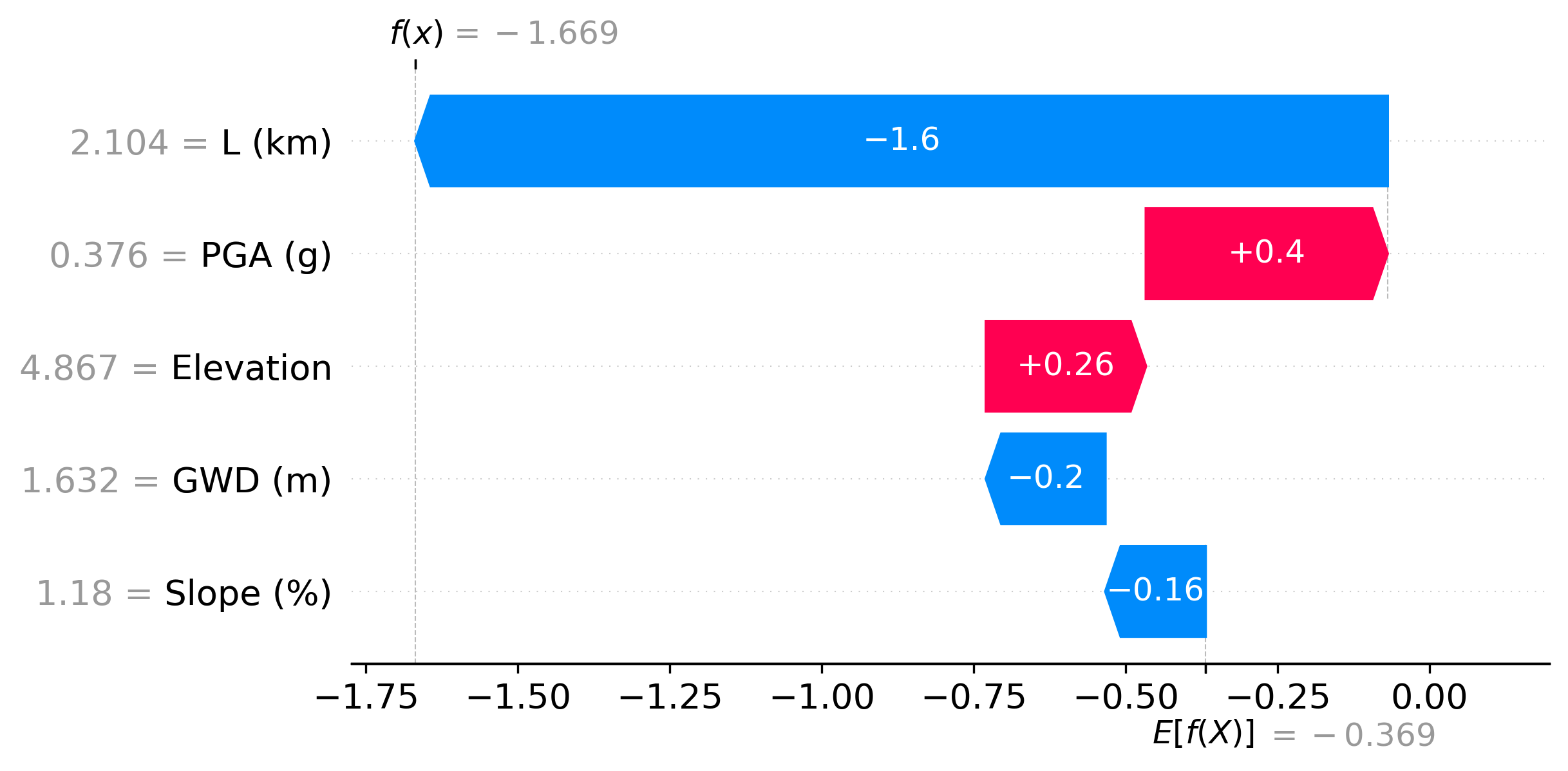}
        \caption{Site 56297 (False negative); 0.388 m displacement; 15.86\%}
        \label{fig:model_a_fn}
    \end{minipage}
\setcounter{figure}{7}
\setcounter{subfigure}{-1}
    \caption{Local explanations of Model A:
    \textbf{(a)} Site 1138 (True positive);
    \textbf{(b)} Site 43457 (True negative);
    \textbf{(c)} Site 32896 (False positive);
    \textbf{(d)} Site 56297 (False negative).
    }
    \label{fig:model_a_localexplanation}
\end{subfigure}
\subsubsection{False positive and false negative}

Next, we look into the false positive in Region B. Site 32896 (see~\Cref{fig:model_a_predictionmap}), with an observed displacement of 0.259 m and labeled as no lateral spread (we use a threshold of 0.3 m to label as lateral spreading), is incorrectly predicted by the XGB model with a 76.60\% probability of lateral spreading occurrence. The SHAP plot (see~\Cref{fig:model_a_fp}) shows that L, PGA, and GWD are the major contributors to the positive total SHAP value of 1.186. The site is near the river (L = 465m), providing a positive SHAP value of 0.62. This value is relatively small compared to the counterpart in~\Cref{fig:model_a_tp} because a larger distance decreases the likelihood of lateral spreading. The SHAP value for the low GWD (1.822 m) is 0.61, while Elevation, PGA, and slope contribute minor effects to the total SHAP values. Although the prediction is incorrect, we argue that the label of no lateral spreading for the site is ambiguous because the observed displacement is close to the defined 0.3 m lateral spreading threshold. Thus, while the prediction is incorrect, the behavior of the site and the SHAP values is aligned with our scientific expectations.

Site 56297 (see~\Cref{fig:model_a_predictionmap}), situated further from the river (L = 2104 m) in Region A, experienced lateral spreading with a displacement of 0.388 m.~\Cref{fig:model_a_fn} shows the local SHAP explanation waterfall plot for this site.  The XGB model predicts no lateral spreading with a 15.86\% probability of occurrence. The SHAP value for L is shown as a negative value of -1.6 due to its large value far from the river. However, we find that the false negatives in Region A align with the Shirley stream (see~\Cref{fig:model_a_predictionmap}), a small stream that was not considered when computing L. Instead, L was computed based on the distance to the larger Avon River, which overestimates L, thus reducing its SHAP value. Recognizing additional water bodies as significant features could enhance the model's predictive accuracy. Nonetheless, lateral spreading near smaller creeks like Dudley Creek and St. Alban Stream are rare, indicating the need for further research to accurately identify influential factors and reduce misclassifications due to data inaccuracies.
Additionally, for this site (ID: 56297), the PGA has the second most prominent effect, a positive SHAP value of +0.4, despite a low PGA (0.367 g). In contrast, the SHAP values for larger PGAs of 0.444 g and 0.473 g (see Figures~\Cref{fig:model_a_tp} and~\Cref{fig:model_a_tn}) are 0.06 and 0.13, significantly smaller than 0.4. Considering only PGAs, we expect the SHAP values to increase with increasing PGAs. The difference in SHAP values with PGAs implies a non-linear and non-monotonic relation between PGA and other factors. To comprehend the effect of various factors and their corresponding SHAP values, we assess them globally.

\subsection{Global explanations}

\Cref{fig:model_a_shapsummary} shows a collective plot of SHAP values of all local explanations from all datasets. Each point represents the SHAP value of a site for a specific feature. The x-axis refers to SHAP values, and the color denotes the magnitude of the feature value (red indicates the feature has a high value, and blue means a low value). The plot highlights how the SHAP value evolves for each feature, which allows us to explain if the model predictions agree with domain knowledge. For example, the SHAP values for L are larger than +2 for L smaller than 0.175 km (i.e., blue points), indicating that sites near the river are predicted to experience lateral spreading. The  SHAP values are smaller than -2 for L greater than 1.71 km (i.e., red points), indicating no lateral spreading at distances far from the river. However, sites that are at a moderate distance from the river are scattered close to the center (i.e., SHAP ~ 0.0), indicating no clear distinction of lateral spreading for medium L ranges, using only L. 

Similarly, SHAP values shift from positive for shallow GWDs (see~\Cref{fig:model_a_shapsummary}), indicating higher lateral spreading likelihood, to negative for deeper GWDs, suggesting lower risk. The SHAP trend for GWD aligns with domain knowledge.~\Cref{fig:model_a_shapsummary} shows that neither elevation nor slope angle relates clearly to SHAP values.~\Cref{fig:model_a_importance} shows the global ranking of each feature based on their mean SHAP value for each site. PGA is the most critical feature among the five input features responsible for a mean SHAP value of 0.84 in all predictions. The elevation feature is the third significant, contributing a mean SHAP value of 0.66 SHAP. In contrast, the slope is the least important feature, contributing an average of 0.19 SHAP value.

While PGA is the most important feature (see~\Cref{fig:model_a_importance} ranking of global features), it does not have a monotonic relationship with SHAP like L and GWD. Contrary to our domain knowledge, we observe high PGAs have large negative SHAP values and a low likelihood of lateral spreading. Also, a few low PGA values have high positive values and a high likelihood of lateral spreading. This contribution of PGA to SHAP value is counterintuitive. 

\begin{subfigure}
\setcounter{figure}{8}
\setcounter{subfigure}{0}
    \centering
    \begin{minipage}[b]{0.45\textwidth}
        \includegraphics[width=\linewidth]{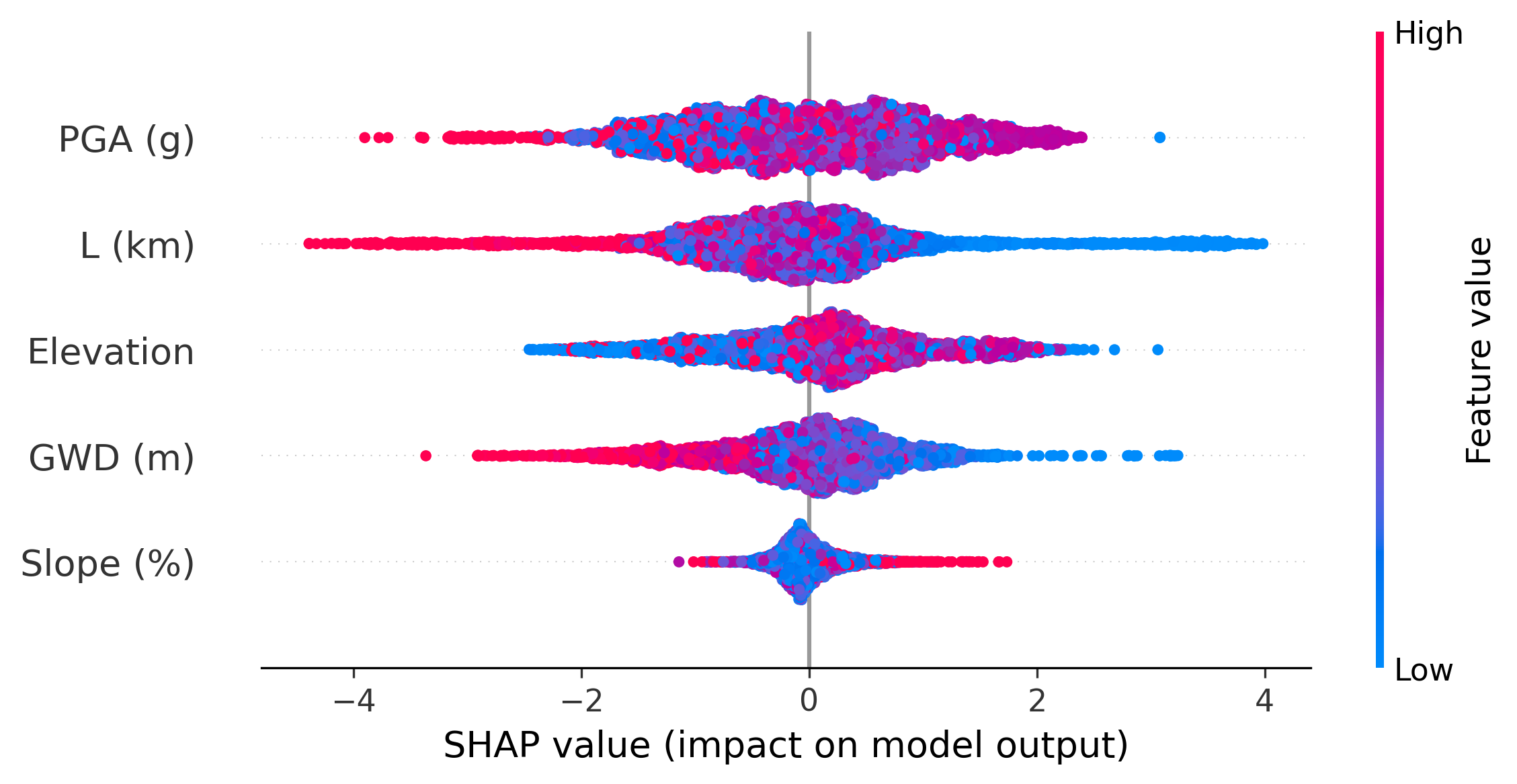}
        \caption{Global explanations}
        \label{fig:model_a_shapsummary}
    \end{minipage}  
\setcounter{figure}{8}
\setcounter{subfigure}{1}
    \begin{minipage}[b]{0.45\textwidth}
        \includegraphics[width=\linewidth]{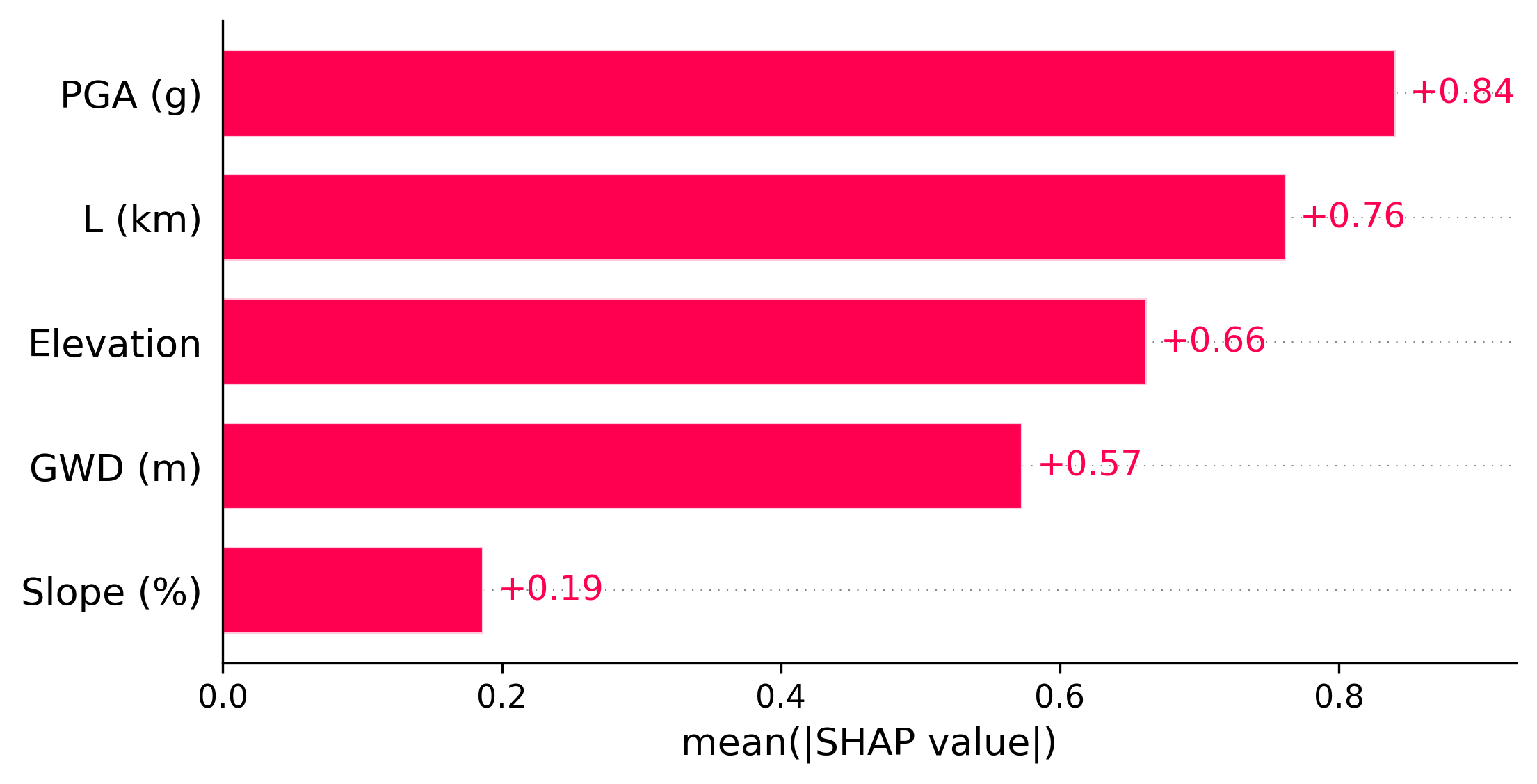}
        \caption{Feature importance}
        \label{fig:model_a_importance}
    \end{minipage}
\setcounter{figure}{8}
\setcounter{subfigure}{-1}
    \caption{Global explanations and feature importance of Model A.
    \textbf{(a)} Global explanations.
    \textbf{(b)} Feature importance.
    }
    \label{fig:model_a_global}
\end{subfigure}
\subsection{Incorrect learning of PGA}

Let us explore how Model A incorrectly learned feature-behavior relation from the dataset using local explanations. We have seen that PGA plays a prominent role in predictive factors. However, examining a TP site prediction (\Cref{fig:model_a_low_pga_shap}), we see that a low PGA of 0.376 g positively influences the SHAP value (+1.29), thus predicting lateral spreading. In contrast, a high PGA of 0.532 g (\Cref{fig:model_a_high_pga_shap}) results in a large negative SHAP value (-2.05), thus a low chance of lateral spreading (22.82\%). Even in these TP and TN categories, the model has incorrectly predicted that high PGAs cause no lateral spreading and that low PGAs are responsible for lateral spreading. 
\begin{subfigure}
\setcounter{figure}{9}
\setcounter{subfigure}{0}
    \centering
    \begin{minipage}[b]{0.45\textwidth}
        \includegraphics[width=\linewidth]{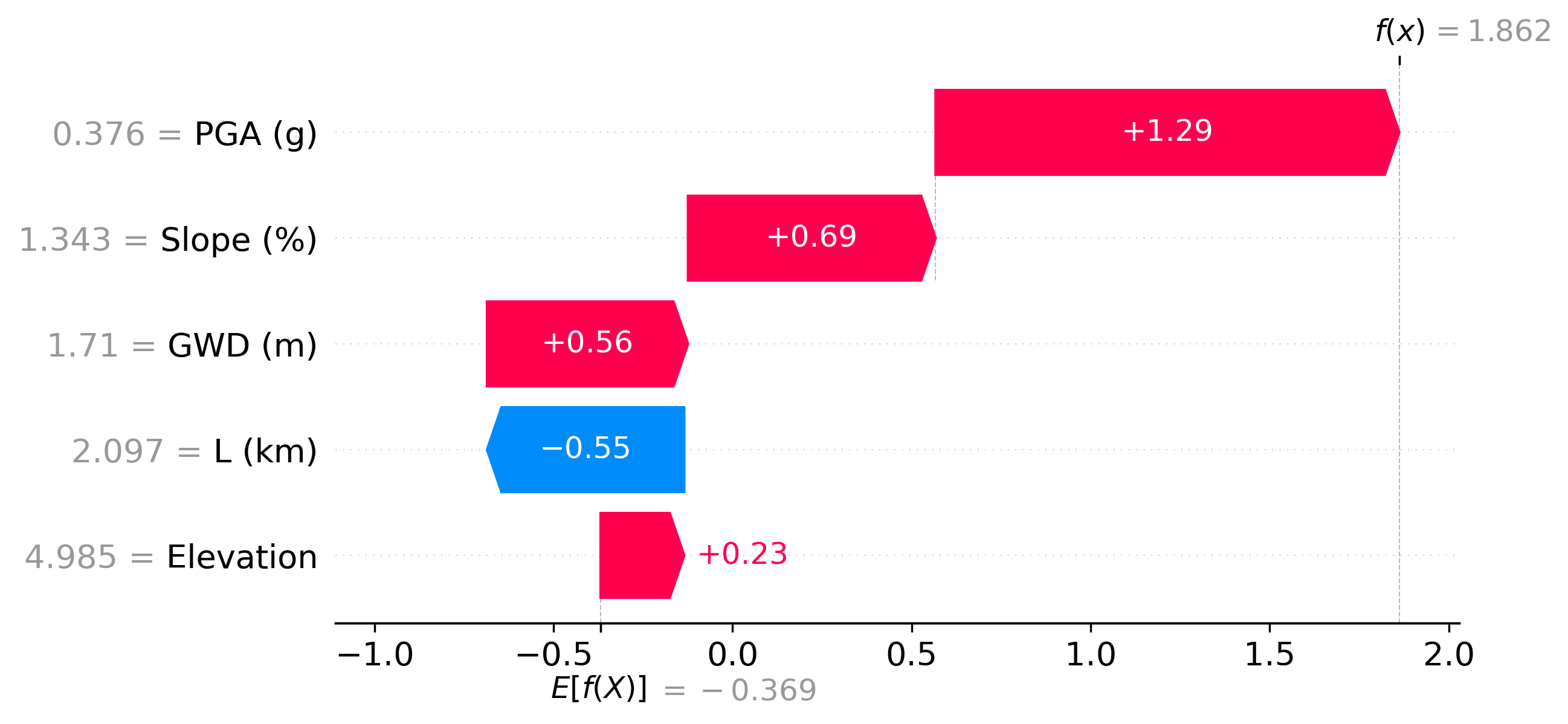}
        \caption{Site 3781 (True positive)}
        \label{fig:model_a_low_pga_shap}
    \end{minipage}  
\setcounter{figure}{9}
\setcounter{subfigure}{1}
    \begin{minipage}[b]{0.45\textwidth}
        \includegraphics[width=\linewidth]{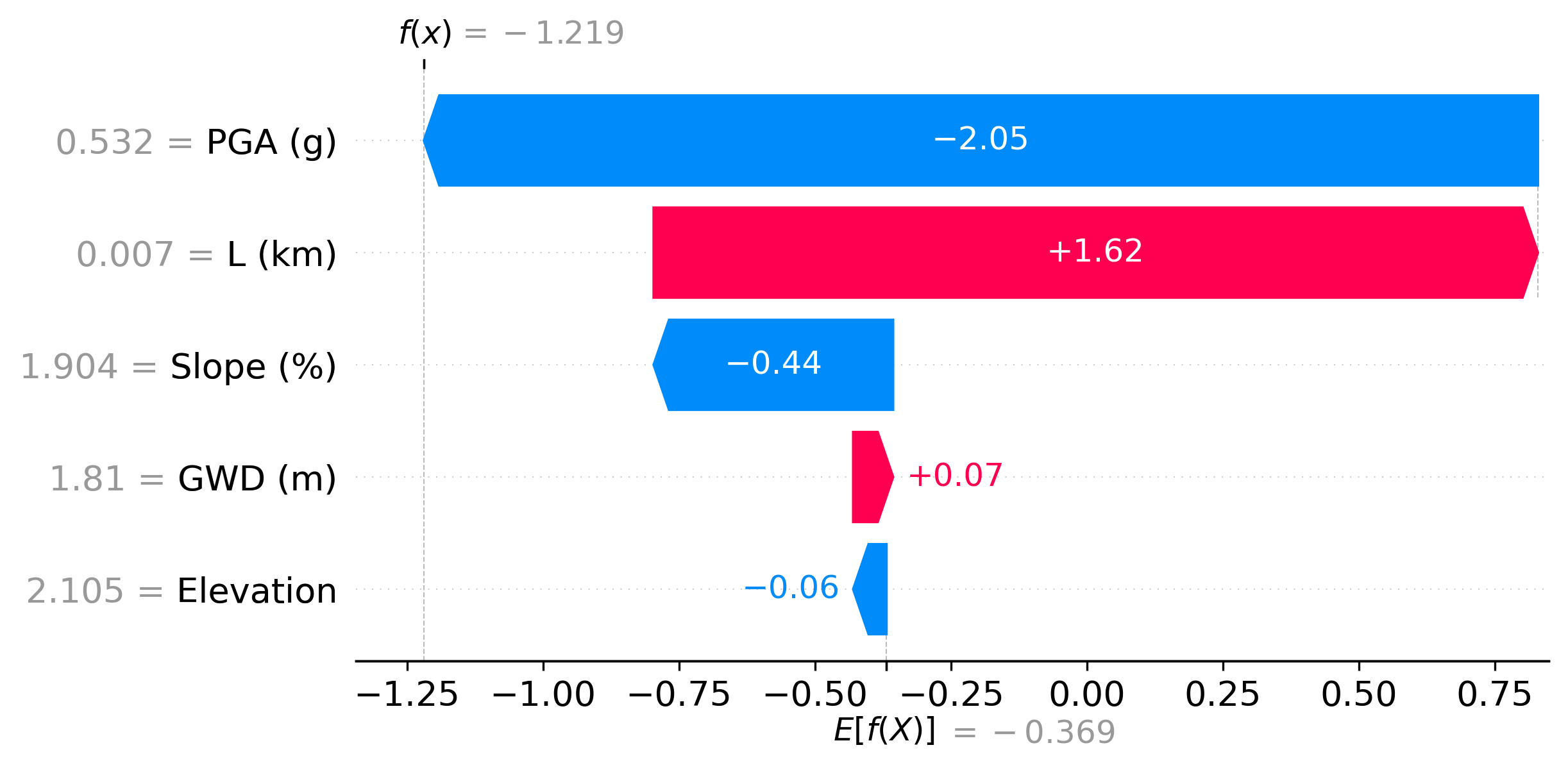}
        \caption{Site 170 (True negative)}
        \label{fig:model_a_high_pga_shap}
    \end{minipage}
\setcounter{figure}{9}
\setcounter{subfigure}{-1}
    \caption{Local waterfall plot explanations for Model A’s incorrect learning of the feature-behavior relation.
    \textbf{(a)} Site 3781
    \textbf{(b)} Site 170
    }
    \label{fig:incorrect_pga_shap}
\end{subfigure}
Of course, this incorrect learning arises from the observations in the training dataset. Inspecting the training set, we see that 46\% of data with PGA in the 0.37 g - 0.375 g experience lateral spreading (48 sites), resulting in positive SHAP values for PGA within this range. On the other hand, 83\% of training data with high PGA ($>0.52 g$) experience no lateral spreading (119 sites), which explains the counterintuitively negative SHAP value for these sites.

As multiple factors influence the prediction for each site, we plot the impact of PGA on SHAP with its strongly correlated elevation feature.~\Cref{fig:shap_pga_el} shows the relationship between SHAP and PGA values. Contrary to expectations, the SHAP values for PGA do not monotonically increase with rising PGA values. Notably, positive SHAP values are concentrated around a PGA of 0.37. Conversely, there is a distinct cluster of negative SHAP values at higher PGA levels ($> 0.52 g$). These patterns corroborate the incorrect model predictions described earlier.

In addition to the relationship between PGA and its SHAP value, Elevation influences the SHAP value (as shown by the color gradient in~\Cref{fig:pga_el_relation}). We observe a clear separation between high (red) and low (blue) elevation values from 0.43 g to 0.47 g of PGA. Within that range of PGA, high elevation values have high SHAP values for PGA, and low elevation values have lower SHAP values for PGA.~\Cref{fig:pga043047el3_map} shows the spatial distribution of lateral spreading observations in training data with PGA ranging from 0.43 g to 0.47 g and with Elevation larger than 3.0 m (corresponding to the purple dots and the red dots within the yellow ellipse in~\Cref{fig:shap_pga_el}).  We observe that 77\% of these sites experience lateral spreading, while only 44\% experience lateral spreading when the elevation is lower than 3.0 m.
\begin{subfigure}
\setcounter{figure}{10}
\setcounter{subfigure}{0}
    \centering
    \begin{minipage}[b]{0.55\textwidth}
        \includegraphics[width=\linewidth]{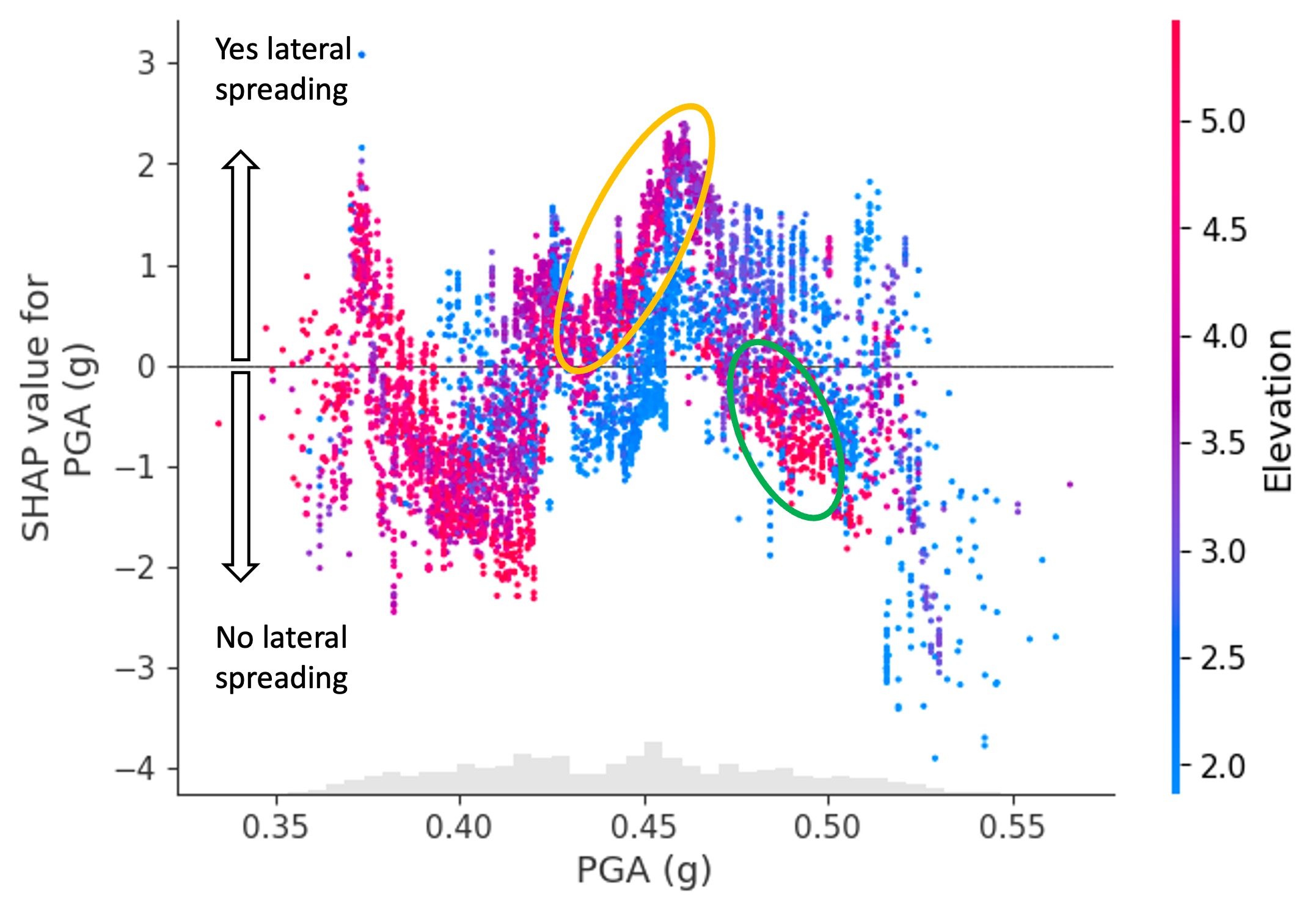}
        \caption{Feature relation between Elevation and PGA and its effect on the SHAP values.}
        \label{fig:shap_pga_el}
    \end{minipage}  
\setcounter{figure}{10}
\setcounter{subfigure}{1}
    \begin{minipage}[b]{0.45\textwidth}
        \includegraphics[width=\linewidth]{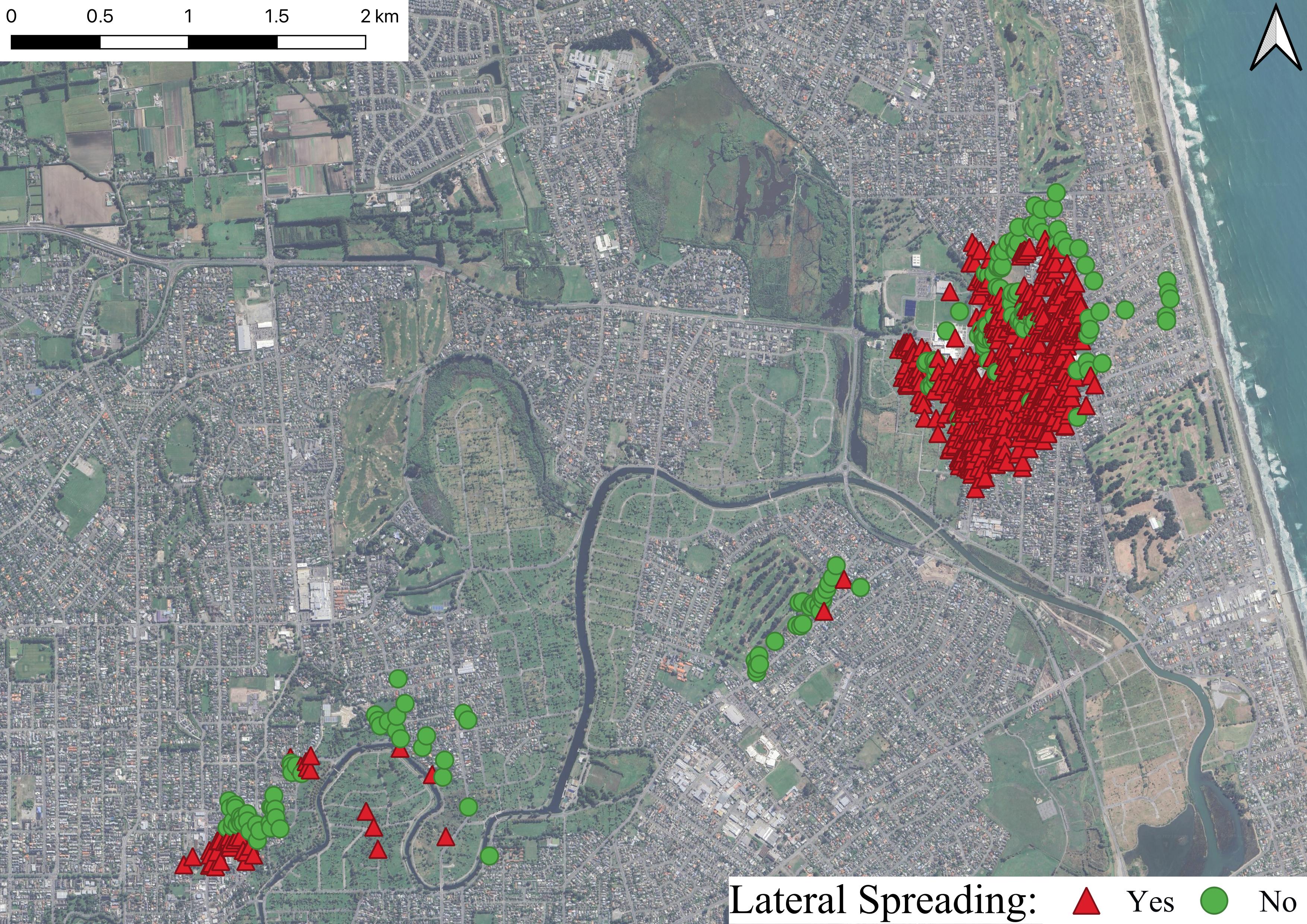}
        \caption{Sites with PGA: 0.43 - 0.47 g and Elevation $>$ 3.0 m}
        \label{fig:pga043047el3_map}
    \end{minipage}
\setcounter{figure}{10}
\setcounter{subfigure}{2}
    \begin{minipage}[b]{0.45\textwidth}
        \includegraphics[width=\linewidth]{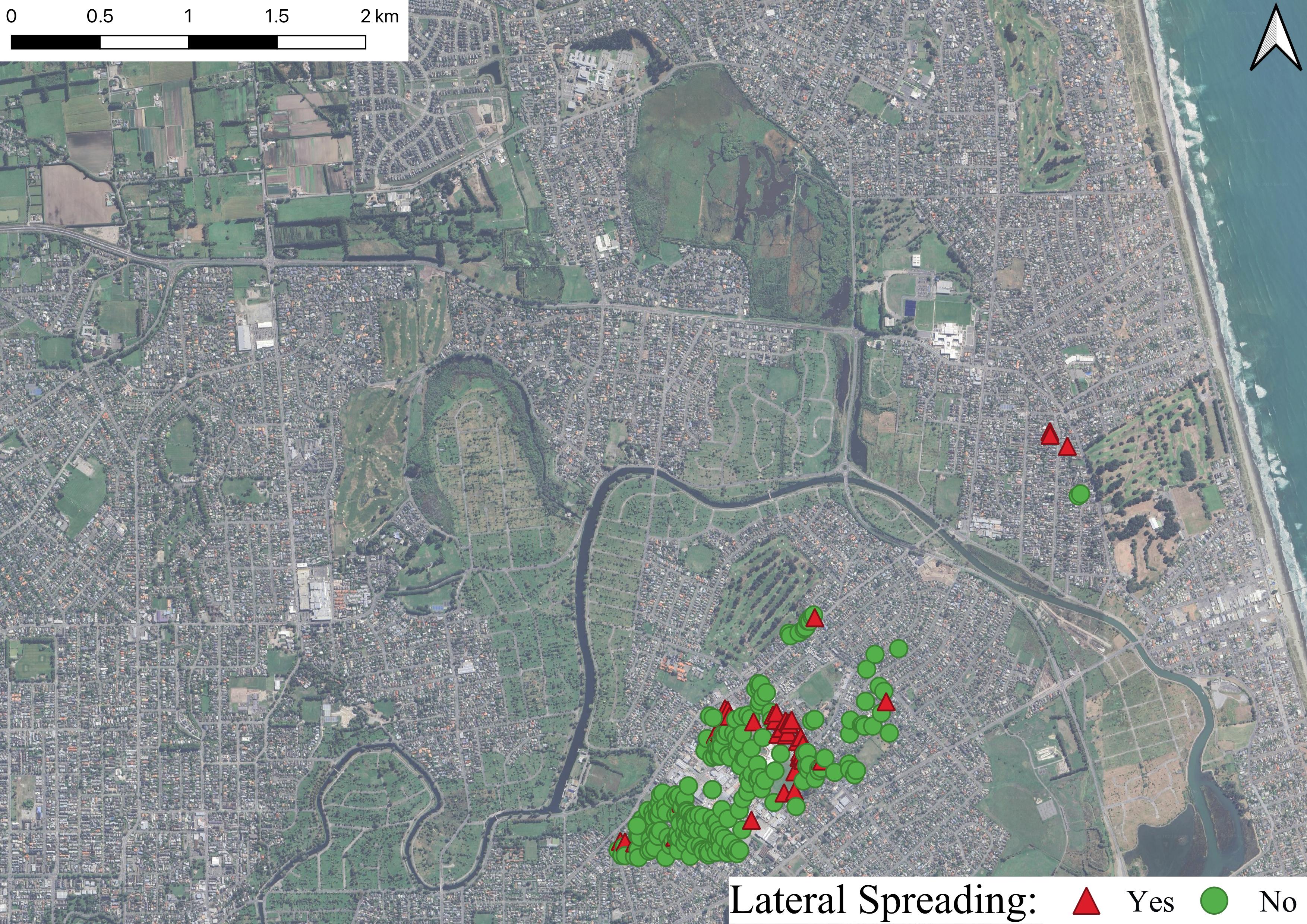}
        \caption{Sites with PGA: 0.47 - 0.50 g and Elevation $>$ 4.0 m}
        \label{fig:pga047050el4_map}
    \end{minipage}
\setcounter{figure}{10}
\setcounter{subfigure}{-1}
    \caption{Feature dependency of PGA and Elevation
    \textbf{(a)} Feature relation between Elevation and PGA and its effect on the SHAP values.
    \textbf{(b)} Sites with PGA: 0.43 - 0.47 g and Elevation $>$ 3.0 m
    \textbf{(c)} Sites with PGA: 0.47 - 0.50 g and Elevation $>$ 4.0 m
    }
    \label{fig:pga_el_relation}
\end{subfigure}
An opposite dependency of elevation is observed in the ranges of 0.47 g to 0.50 g of PGA, where high elevation values have negative SHAP values and low elevation values have positive SHAP values.~\Cref{fig:pga047050el4_map} shows the observations with PGA ranging from 0.47 g to 0.50 g and elevation larger than 4.0 m (corresponding to the red dots within the green ellipse in~\Cref{fig:shap_pga_el}). 84\% of data experience no lateral spreading, while only 42\% experience no lateral spreading when elevation is below 4.0 m. These observations agree with the separation observed in~\Cref{fig:shap_pga_el}, meaning elevation impacts the SHAP values for PGA.

\Cref{fig:shap_pga_el} shows a non-linear relation between PGA and SHAP values. Low PGAs (0.35 - 0.4 g) show no lateral spreading, while PGAs in the 0.43 - 0.47 g range show a higher chance of lateral spreading (positive SHAP values), and as PGA increases above 0.47 g, the sites show no signs of lateral spreading. The chance of lateral spreading decreasing with PGA is counterintuitive.  Consider the region within the yellow circle in ~\Cref{fig:shap_pga_el}, which shows positive SHAP values for PGAs between 0.43 - 0.47 g and elevation larger than 3 m. This cluster of lateral spreading occurrences (sites marked within the yellow circle in~\Cref{fig:shap_pga_el}) is shown on a map in~\Cref{fig:pga043047el3_map}. While this cluster shows signs of lateral spreading based on optical image correlation (\Cref{fig:displ-map}), these sites do not exhibit traditional signs of lateral spreading, such as observation of cracks, and disagree with the observations (\Cref{fig:nzgd-map}) recorded by the New Zealand Geotechnical Database (NZGD,~\citeyear{nzgd-2013}). The manifestations recorded by NZGD in this area are categorized as ``minor liquefaction with no apparent lateral movement." This disparity arises from differences in the methods used to measure lateral movement. NZGD's measurements are based on the displacement observed based on cracks, whereas the lateral spreading occurrences documented in this study rely on optical image correlation of horizontal displacement. This difference explains the reasoning behind positive SHAP values for PGA between 0.43 - 0.47 g with elevation larger than 3.0 m.

Let us explore the reasoning behind no lateral spreading cases for high PGAs ($> 0.47 g$), marked by a green circle in~\Cref{fig:shap_pga_el}. The region of interest is mapped in~\Cref{fig:pga047050el4_map} and experiences significant no lateral spreading. Although this region experiences higher PGA, it is situated at a much higher elevation and located further away from the river, thus showing no signs of lateral spreading. 

\subsubsection{Effect of CPT on Lateral Spreading Predictions}
\begin{figure}
\begin{center}
\includegraphics[width=12cm]{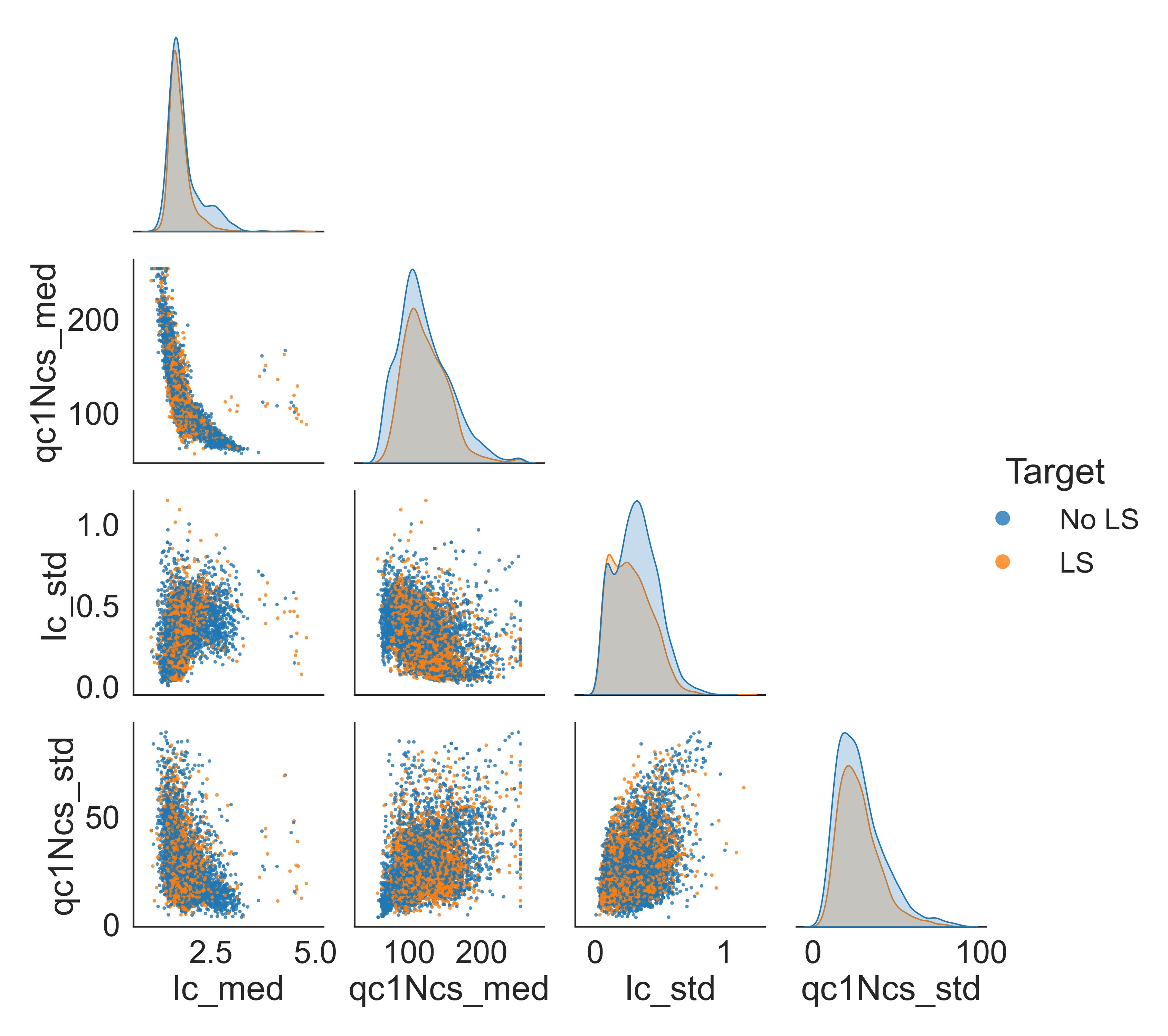}
\end{center}
\caption{Pairplots of CPT features with lateral spreading observations.}
\label{fig:cpt_pairplots}
\end{figure}

In addition to topological features, the likelihood of lateral spreading also depends on the site characteristics – i.e., its soil properties. Therefore, we introduce CPT features to account for the soil properties of the site. This study uses the same CPT features as~\citet{durante-2021}: the median and the standard deviation of $q_{c1Ncs}$ and $I_c$ over 4 m below the groundwater table. The parameters $q_{c1Ncs}$ and $I_c$ represent the soil strength and type. A site with high $q_{c1Ncs}$ or $I_c$ means the soil is strong or clayey to liquefy and, consequently, has a low risk of lateral spreading. The medians and standard deviations represent the profile's general trend and variance over 4 m below the GWD.~\Cref{fig:cpt_pairplots} shows the pair plots of the four CPT features—the plots on the diagonal show the histogram of each feature. The plots in the lower triangle refer to the scatter plot of each pair of features. We observe that the median of $q_{c1Ncs}$ and $I_c$ have a strong negative correlation, while other pairs of features do not show strong correlations. In~\Cref{fig:cpt_pairplots}, blue means no lateral spreading, and orange represents lateral spreading. 
We notice a higher frequency of no lateral spreading cases for sites with fine-grained soil ($I_{c\_med}$ value exceeds 2.5). Fine-grained soils are less prone to generating excess pore water pressure during dynamic loading, making it difficult to liquefy or experience lateral spreading. Similarly, when $q_{c1Ncs}$ med surpasses 175, we also observe a greater prevalence of no lateral spreading cases. High $q_{c1Ncs}$ med values indicate dense soil, which offers higher shear strength against dynamic loading, reducing the likelihood of lateral spreading. Despite the expectation that regions with low $q_{c1Ncs}$ med ($<75$), i.e., loose soil, are more susceptible to liquefaction, we find a clustering of cases without lateral spreading in this region. However, most low $q_{c1Ncs}$ med sites in our study region have high $I_{c\_med}$ values (see red circle in~\Cref{fig:cpt_pairplots}), indicating that the loose soils are fine-grained and less susceptible to liquefaction.  Hence it is important to correlate CPT features to lateral spreading to improve our model predictions.

We develop another XGBoost model, Model B, using the same dataset as Model A with four additional CPT features as input parameters (see~\Cref{tab:model_info}). After optimization, all hyperparameters are the same, except Model B uses only 26 estimators, as the validation performance does not improve with more estimators. Model B yields an accuracy of 86.5\% on the validation dataset and 82.77\% on the testing dataset.~\Cref{tab:model_info} compares the performance of Models A and B. Adding CPT features did not improve the performance of Model B compared to Model A. This result agrees with~\citet{durante-2021}. We will dive into local and global explanations of Model B to evaluate why adding CPT features did not improve the model's performance.

\subsubsection*{Global Explanation of Model B:}

To evaluate the global performance of Model B, we summarize all the local explanations of Model B predictions in~\Cref{fig:model_b_shapsummary}. The trend of SHAP values for Model B aligns with Model A. Both GWD and L show behavior that aligns with our expectations, and the SHAP values of PGA show a counterintuitive trend. However,  the CPT features do not show a clear trend.
\begin{subfigure}
\setcounter{figure}{12}
\setcounter{subfigure}{0}
    \centering
    \begin{minipage}[b]{0.45\textwidth}
        \includegraphics[width=\linewidth]{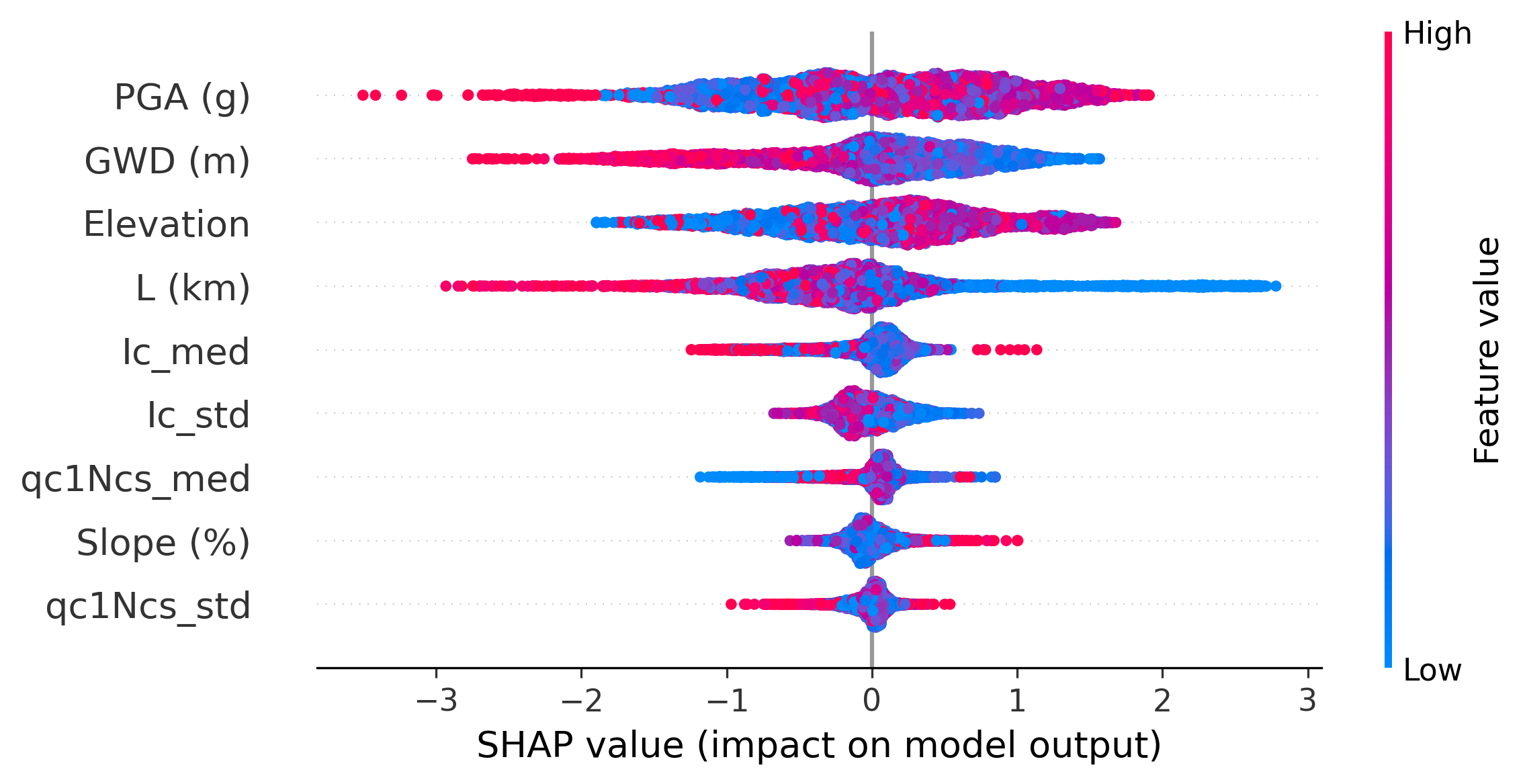}
        \caption{Global explanation}
        \label{fig:model_b_shapsummary}
    \end{minipage}  
\setcounter{figure}{12}
\setcounter{subfigure}{1}
    \begin{minipage}[b]{0.45\textwidth}
        \includegraphics[width=\linewidth]{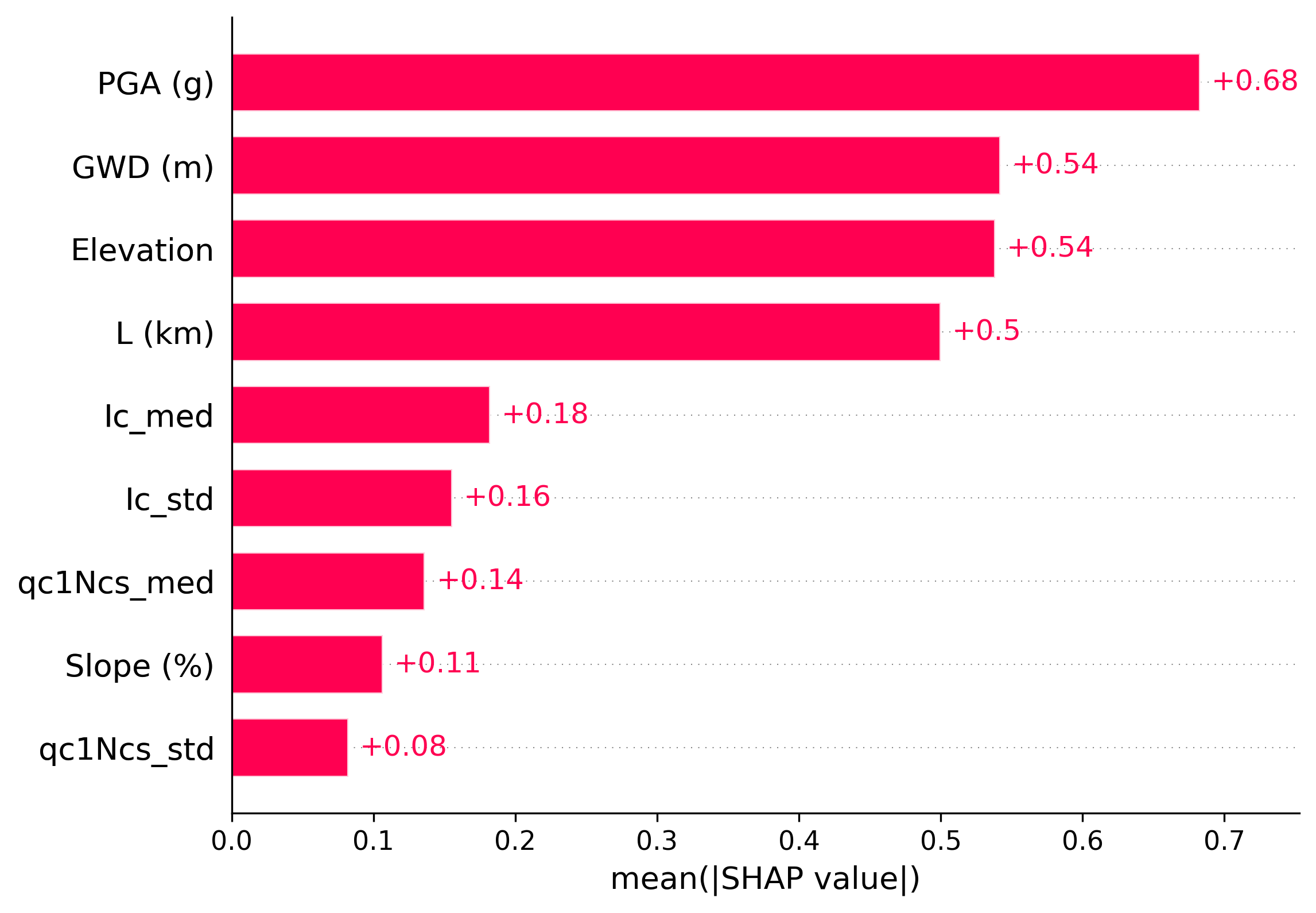}
        \caption{Feature importance}
        \label{fig:model_b_importance}
    \end{minipage}
\setcounter{figure}{12}
\setcounter{subfigure}{-1}
    \caption{Global explanations and feature importance of Model B
    \textbf{(a)} Global explanation
    \textbf{(b)} Feature importance
    }
    \label{fig:model_b_global}
\end{subfigure}
Even after adding site-specific soil properties through CPT,~\Cref{fig:model_b_importance} shows that PGA remains the most influential factor in Model B, with a mean SHAP value of 0.68. GWD and Elevation exhibit similar contributions to predictions, each with a mean SHAP value of 0.54. L becomes less critical in Model B than in Model A, ranking fourth. On the other hand, including all four CPT features only make minor contributions to predictions.

\subsubsection*{Local Explanation of Model B:}

Let us explore the local explanations of Model B to understand the impact of CPT features. Site 5252 (see~\Cref{fig:model_a_predictionmap}), situated near the river (L = 99 m) with a moderate PGA of 0.42 g and a deep GWD of 2.569 m, experiences lateral spreading. Model A accurately predicts this occurrence with a 68.46\% probability. The SHAP waterfall plot for Model A (\Cref{fig:model_a_5252}) highlights the proximity to the river, contributing a decisive SHAP value of 1.414, leading to a true positive prediction. However, including CPT features results in an incorrect prediction of no-lateral-spreading by Model B (\Cref{fig:model_b_5252}). Despite L contributing a positive SHAP value of 0.59, the median values of $I_c$ (2.546) and $q_{c1Ncs}$ (69.031) counteract its effect with SHAP values of -0.8 and -0.55, respectively. High $I_c$ and low $q_{c1Ncs}$ indicate that clayey soil is less prone to liquefaction and lateral spreading. While the negative SHAP values align with our understanding of clayey soil, they result in a false negative prediction.
\begin{subfigure}
\setcounter{figure}{13}
\setcounter{subfigure}{0}
    \centering
    \begin{minipage}[b]{0.45\textwidth}
        \includegraphics[width=\linewidth]{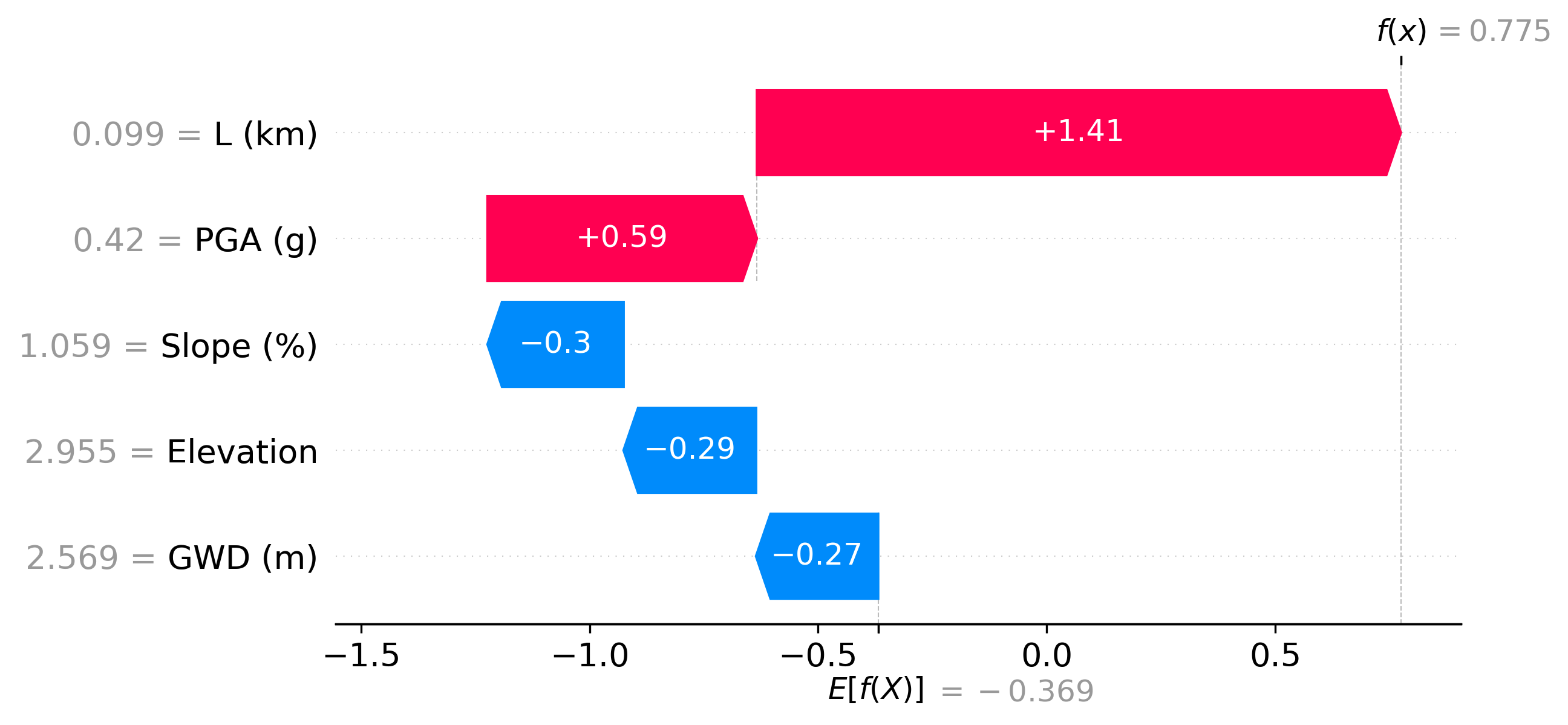}
        \caption{Site 5252 in Model A (True positive)}
        \label{fig:model_a_5252}
    \end{minipage}  
\setcounter{figure}{13}
\setcounter{subfigure}{1}
    \begin{minipage}[b]{0.45\textwidth}
        \includegraphics[width=\linewidth]{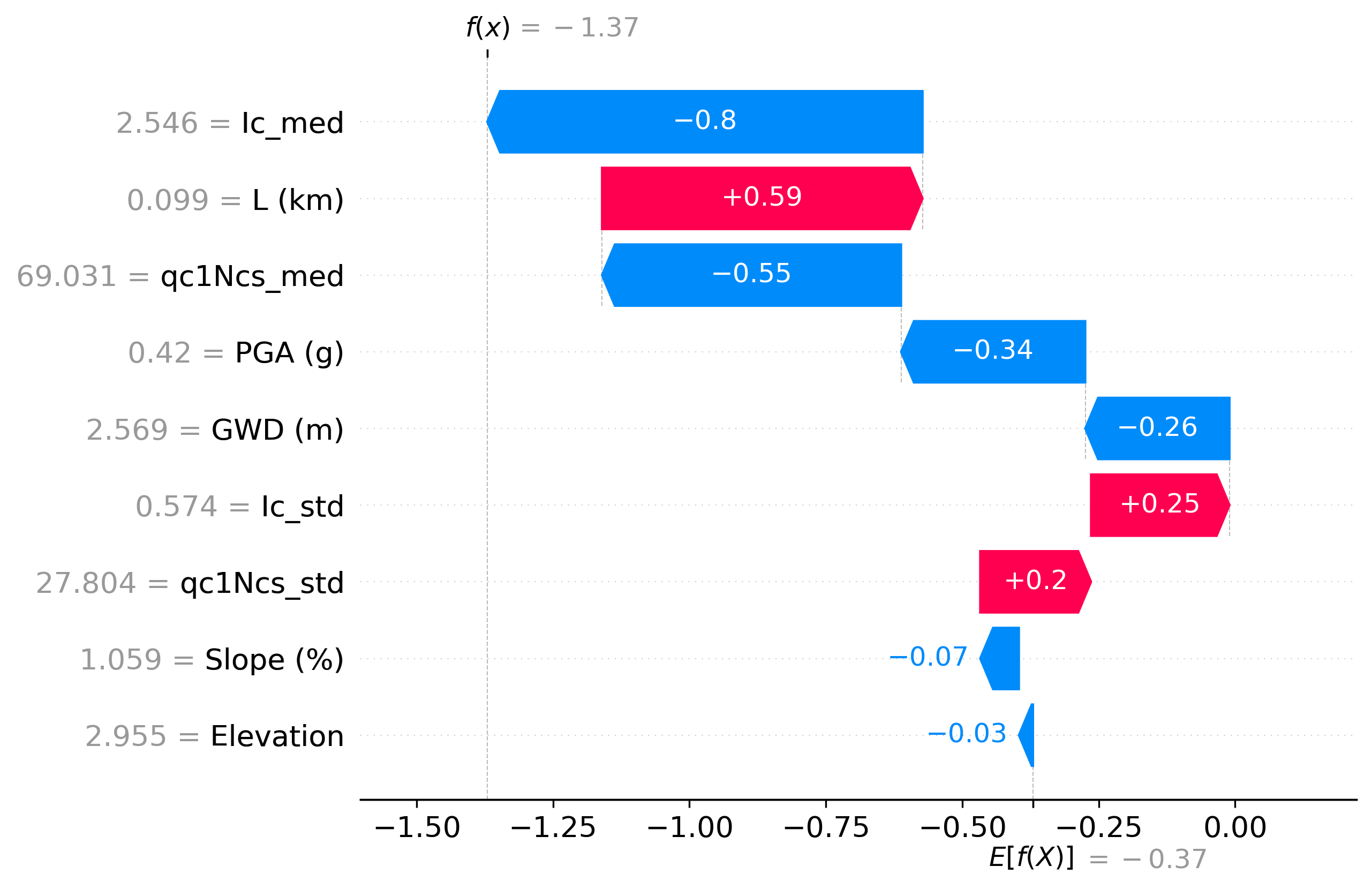}
        \caption{Site 5252 in Model B (False negative)}
        \label{fig:model_b_5252}
    \end{minipage}
\setcounter{figure}{13}
\setcounter{subfigure}{2}
    \begin{minipage}[b]{0.45\textwidth}
        \includegraphics[width=\linewidth]{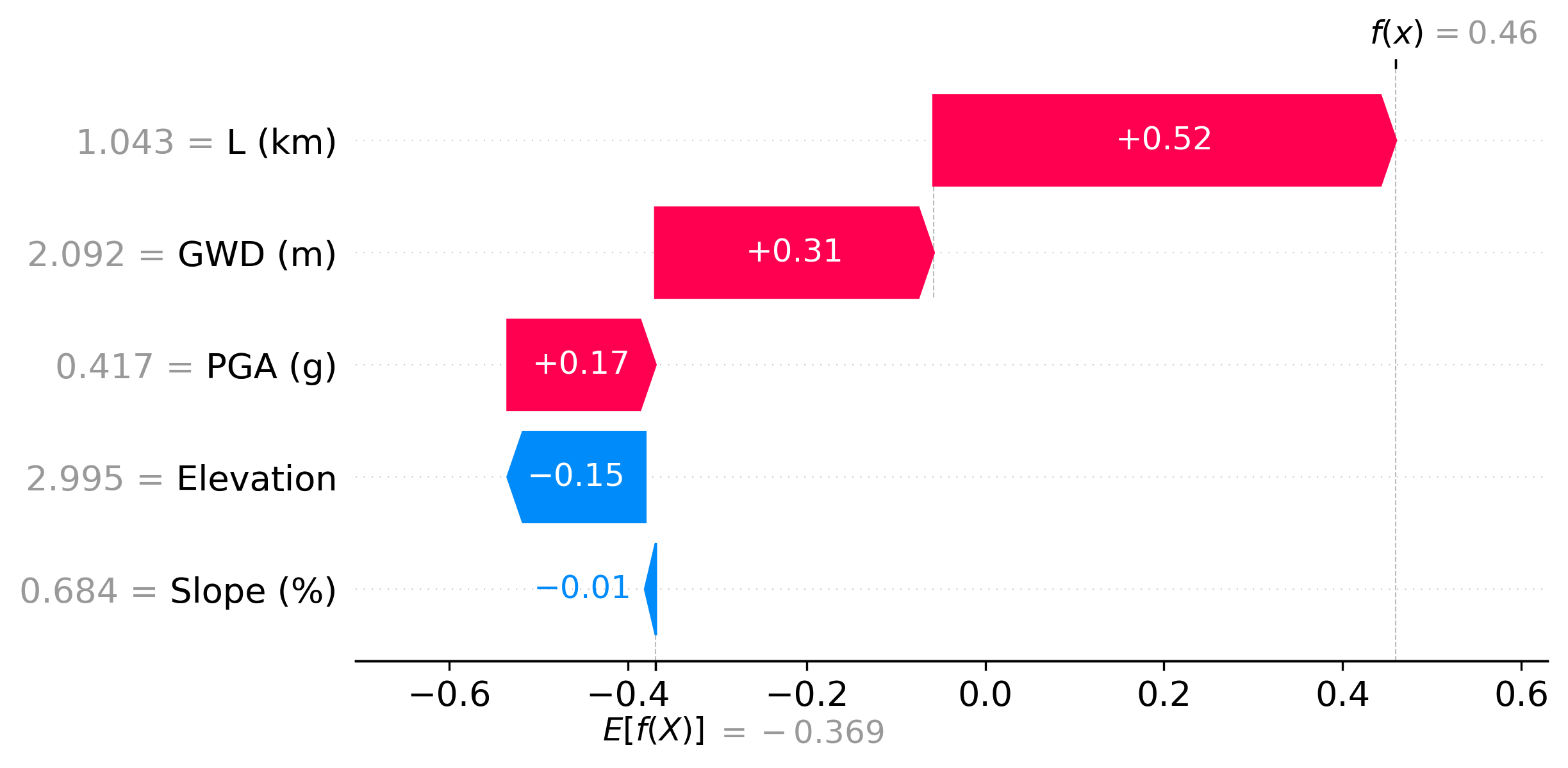}
        \caption{Site 17049 in Model A (False positive)}
        \label{fig:model_a_17049}
    \end{minipage}
\setcounter{figure}{13}
\setcounter{subfigure}{3}
    \begin{minipage}[b]{0.45\textwidth}
        \includegraphics[width=\linewidth]{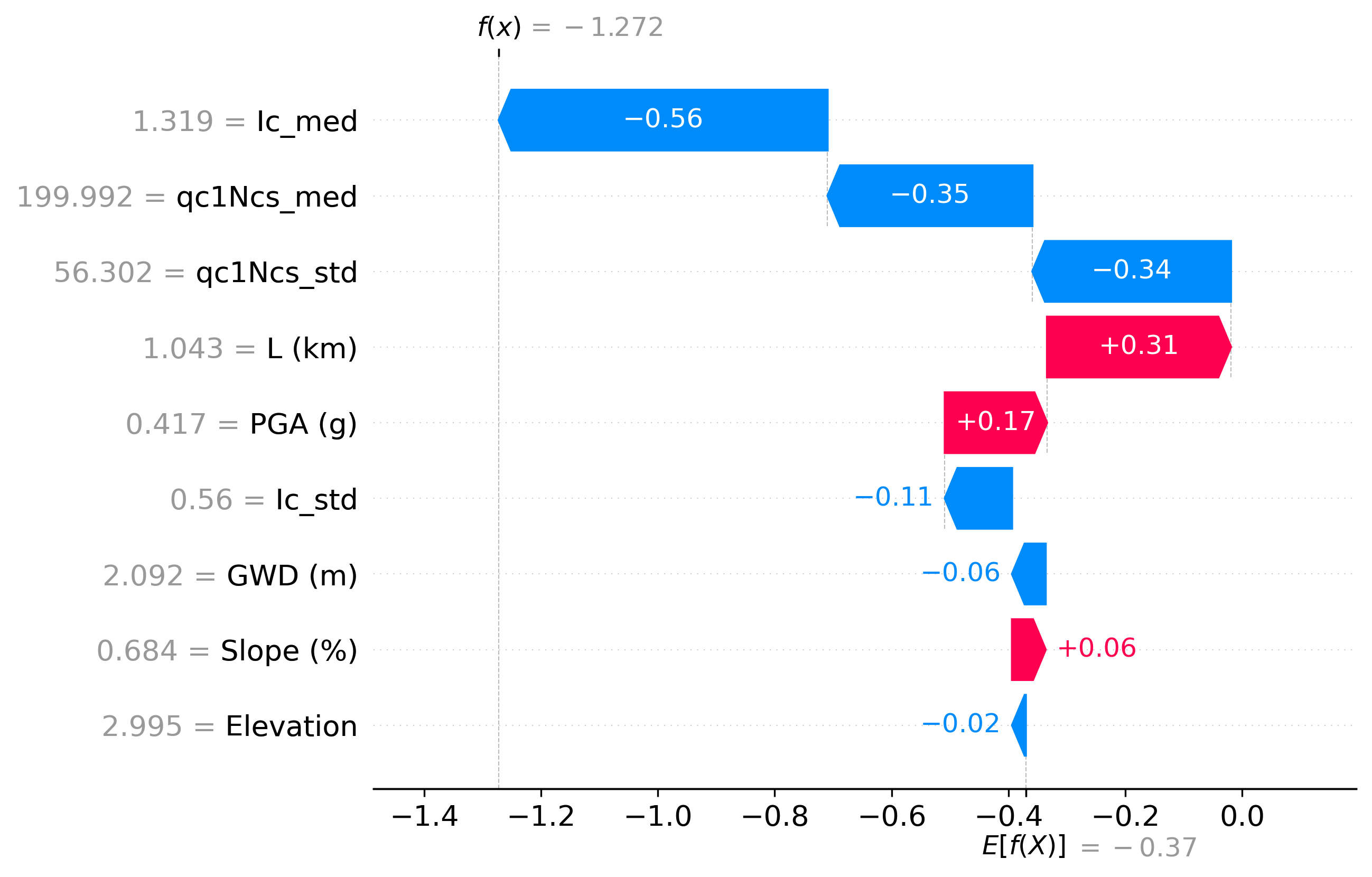}
        \caption{Site 17049 in Model B (True negative)}
        \label{fig:model_b_17049}
    \end{minipage}
\setcounter{figure}{13}
\setcounter{subfigure}{-1}
    \caption{Local explanation of Site 5252 and Site 17049
    \textbf{(a)} Site 5252 in Model A (True positive);
    \textbf{(b)} Site 5252 in Model B (False negative);
    \textbf{(c)} Site 17049 in Model A (False positive);
    \textbf{(d)} Site 17049 in Model B (True negative).
    }
    \label{fig:local_comparison}
\end{subfigure}
In contrast, Site 17049 (see~\Cref{fig:model_a_predictionmap}) benefits from considering the CPT features. Model A misclassifies the site as prone to lateral spreading, but Model B accurately predicts lateral spreading by including CPT data. The site is located 1 km from the Avon River with a GWD of 2 m, and Model A incorrectly predicted a 61\% chance of lateral spreading due to positive SHAP values in both L and GWD of 0.52 and 0.31 (see~\Cref{fig:model_a_17049}). However, Model B (see~\Cref{fig:model_b_17049}), incorporating CPT features like the low $I_c$  (1.319) and high  $q_{c1Ncs}$ (199.992), yields negative SHAP values of -0.56 and -0.35, indicating strong sandy soil resistant to liquefaction, leading to a correct prediction of no lateral spreading.

\subsubsection*{What Model B has learned from CPT features:}

While incorporating CPT features does not notably enhance predictive performance, understanding Model B's interpretation of these features is interesting.~\Cref{fig:icqc_shap_comparison} presents the SHAP values of features plotted in median $I_c$ and median $q_{c1Ncs}$ space, where the color indicates the sign and magnitude of the SHAP value.

\Cref{fig:ic_shap} illustrates the SHAP value of median $I_c$. A distinct separation between positive (red) and negative (blue) SHAP values is noticeable at $I_c=1.5$ and $I_c=2.4$. When $I_c$ is less than 1.5, indicating coarse-grained/sandy soil, the SHAP value ranges from 0 to -0.5, while for $I_c$ greater than 2.4, indicating fine-grained soil, the SHAP value ranges from -0.5 to -1.0. These negative SHAP values represent no lateral spreading, which is expected for fine-grained soil but not sandy soil. However, these small  $I_c$ values are associated with large $q_{c1Ncs}$, so the soil does not liquefy, and lateral spreading does not occur.   In the intermediate range ($1.5<I_c<2.4$), the SHAP value ranges from -0.1 to 0.3.  For these silty/sandy soils, the potential for lateral spreading is not directly associated with  $I_c$, but rather with the relative density of the soil as indicated by $q_{c1Ncs}$. 
\begin{subfigure}
\setcounter{figure}{14}
\setcounter{subfigure}{0}
    \centering
    \begin{minipage}[b]{0.3\textwidth}
        \includegraphics[width=\linewidth]{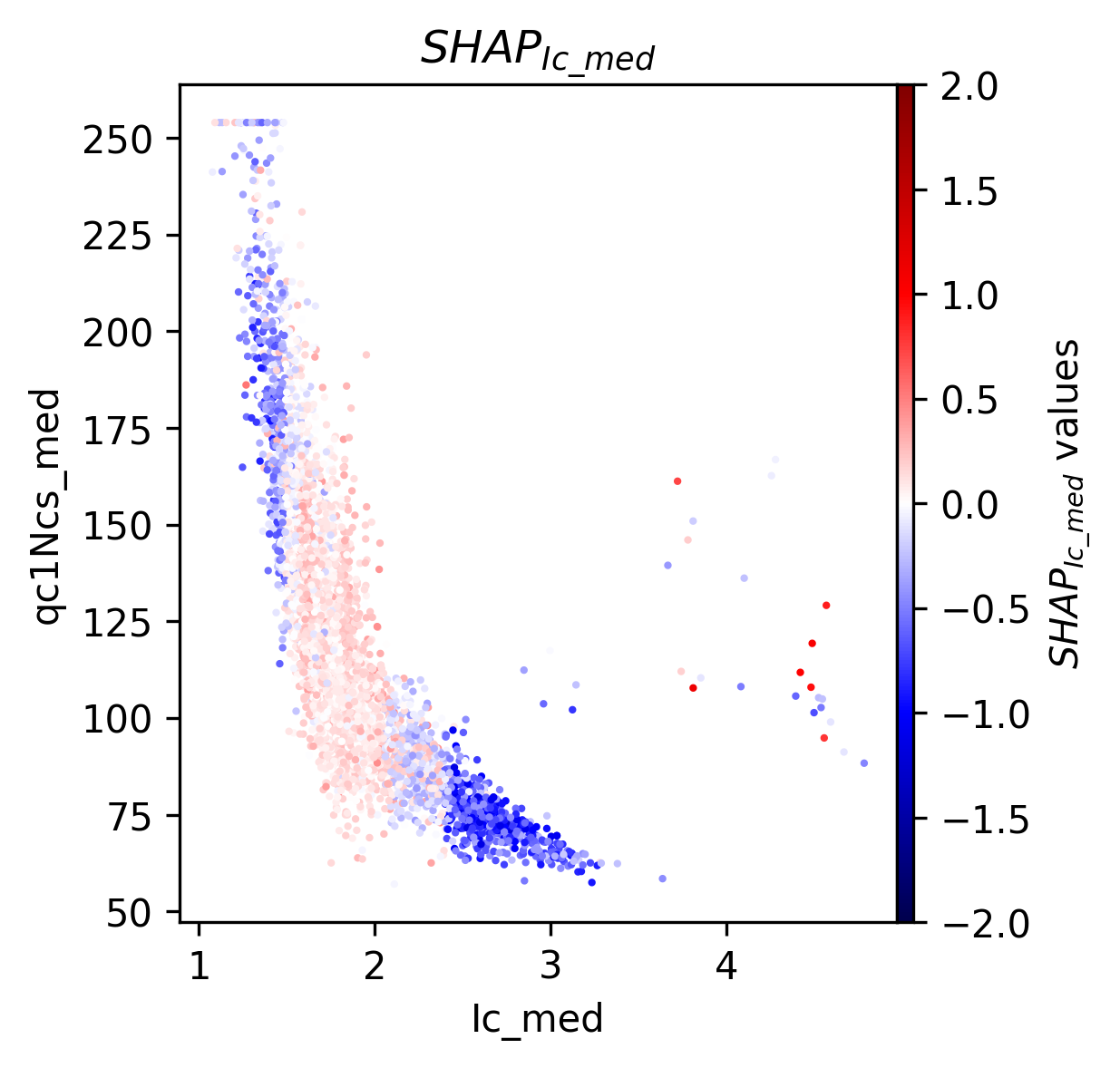}
        \caption{SHAP values for $I_c$ median}
        \label{fig:ic_shap}
    \end{minipage}  
\setcounter{figure}{14}
\setcounter{subfigure}{1}
    \begin{minipage}[b]{0.3\textwidth}
        \includegraphics[width=\linewidth]{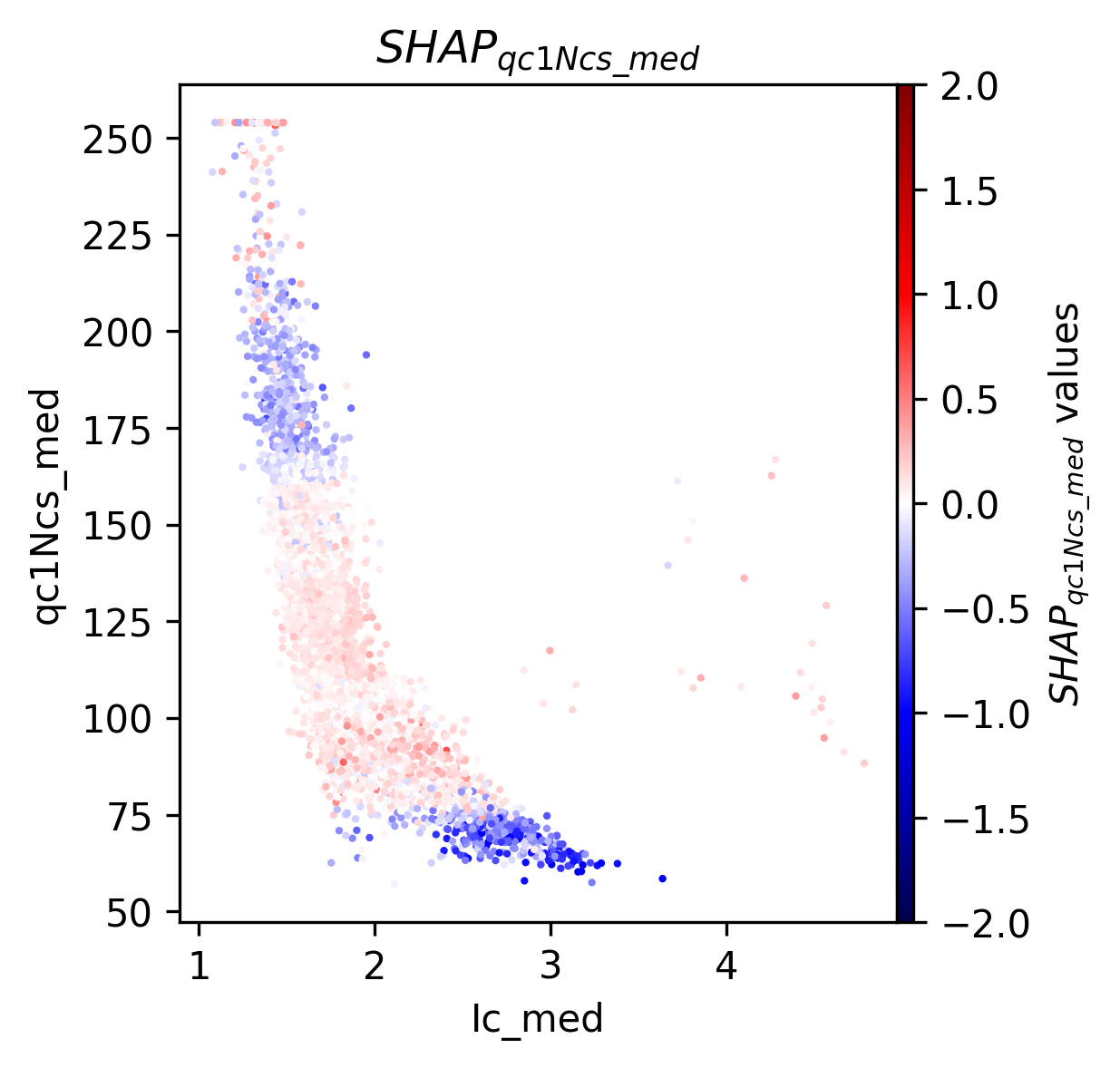}
        \caption{SHAP values for $q_{c1Ncs}$ median}
        \label{fig:qc_shap}
    \end{minipage}
\setcounter{figure}{14}
\setcounter{subfigure}{2}
    \begin{minipage}[b]{0.3\textwidth}
        \includegraphics[width=\linewidth]{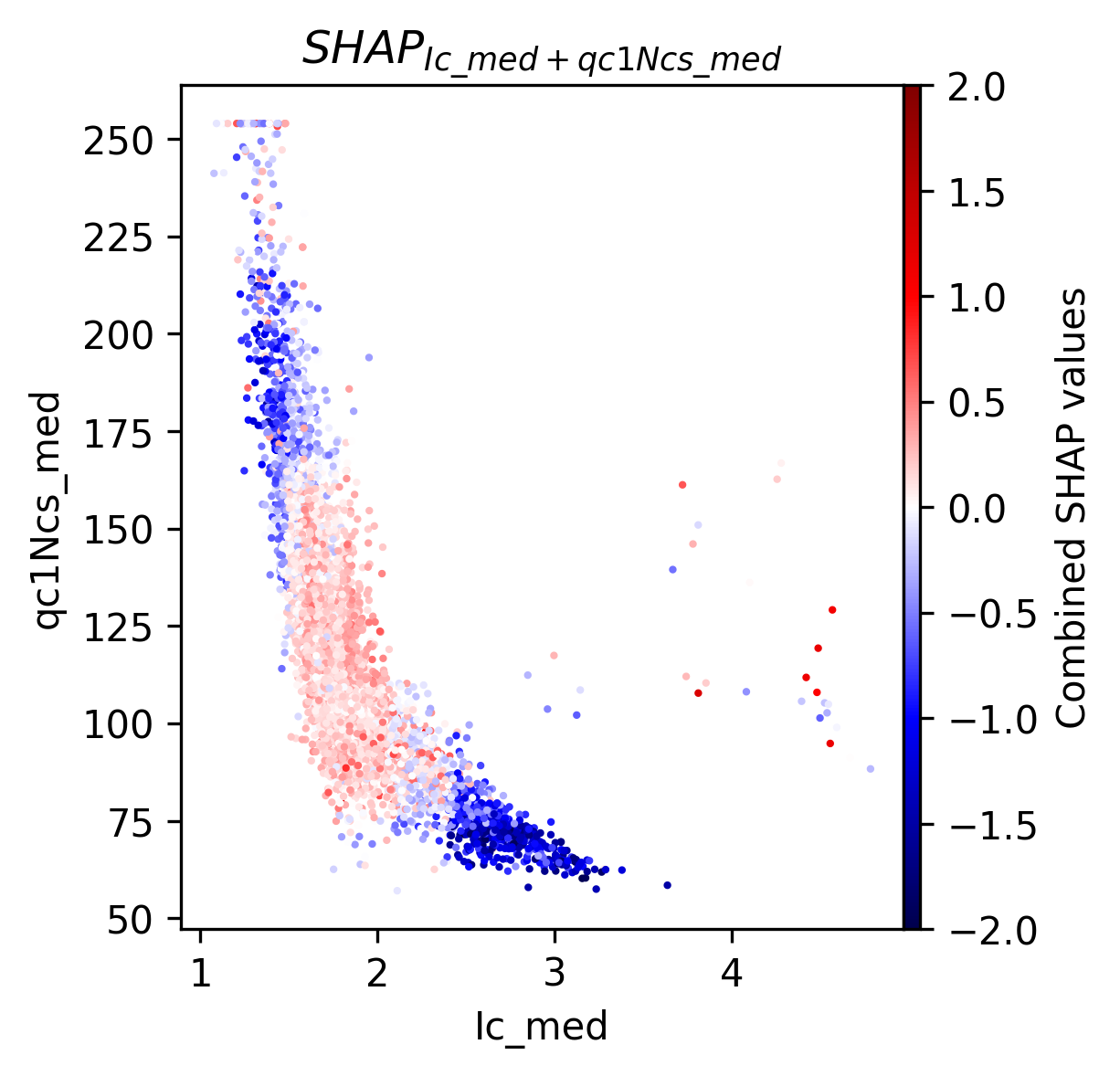}
        \caption{Combined SHAP values}
        \label{fig:combined_shap}
    \end{minipage}
\setcounter{figure}{14}
\setcounter{subfigure}{-1}
    \caption{SHAP values for medians of Ic and qc1Ncs in Model B
    \textbf{(a)} SHAP values for $I_c$ median;
    \textbf{(b)} SHAP values for $q_{c1Ncs}$ median;
    \textbf{(c)} Combined SHAP values of $I_c$ median and $q_{c1Ncs}$ median.
    }
    \label{fig:icqc_shap_comparison}
\end{subfigure}
A similar pattern appears in the SHAP values for median $q_{c1Ncs}$ (\Cref{fig:qc_shap}). When the median $q_{c1Ncs}$ falls between 160 and 210, the SHAP value ranges from -0.5 to 0, indicating no lateral spreading due to the dense nature of the soil. For $q_{c1Ncs}$ values below 75, the SHAP value ranges from 0 to -1.0, indicating no lateral spreading due to the large $I_c$ associated with these small values of $q_{c1Ncs}$. For $q_{c1Ncs}$ between 75 and 160, the SHAP value is approximately 0 to 0.2.

Comparing ~\Cref{fig:ic_shap,fig:qc_shap}, we observe a similar trend in SHAP values for median $I_c$ and $q_{c1Ncs}$. Low median $I_c$ data often corresponds to high median $q_{c1Ncs}$, and vice versa, resulting in a similar trend of SHAP values.~\Cref{fig:combined_shap} combines the SHAP values for median $I_c$ and $q_{c1Ncs}$, revealing regions of positive (red) and negative (blue) SHAP values consistent with our understanding of liquefaction and soil properties, as explained earlier in this section. 

\subsubsection*{Improving model performance with CPT data:}
Although Model B effectively learns the relationship between SHAP values for $I_c$ and $q_{c1Ncs}$, their incorporation does not improve predictive performance. Additionally,~\Cref{fig:model_b_importance} indicates that the CPT features ($q_{c1Ncs}$ ) are least important for the model predictions. Hence, we develop a new Model C by excluding the three least important features identified in~\Cref{fig:model_b_importance}: the median and standard deviation of $q_{c1Ncs}$, and slope. Model C exhibits improved performance, achieving accuracy rates of 88.04\% and 85.54\% on the validation and testing data, which are higher than Model B (testing accuracy is 82.77\%). Since $q_{c1Ncs}$ and $I_c$ are negatively correlated, including only the $I_c$ factors improves the model's performance. 
\begin{subfigure}
\setcounter{figure}{15}
\setcounter{subfigure}{0}
    \centering
    \begin{minipage}[b]{0.45\textwidth}
        \includegraphics[width=\linewidth]{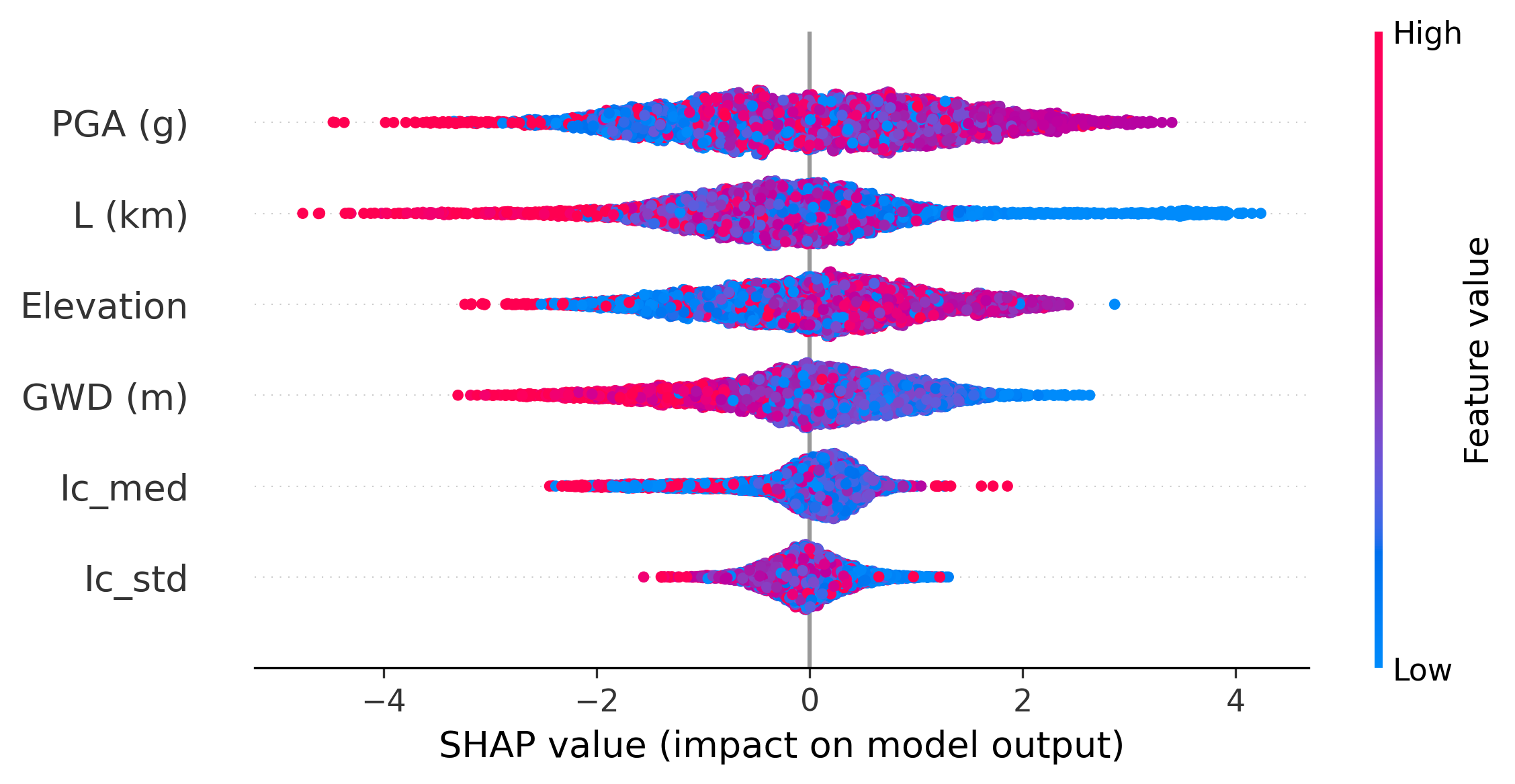}
        \caption{Global explanation}
        \label{fig:shap_summary_modelc}
    \end{minipage}  
\setcounter{figure}{15}
\setcounter{subfigure}{1}
    \begin{minipage}[b]{0.45\textwidth}
        \includegraphics[width=\linewidth]{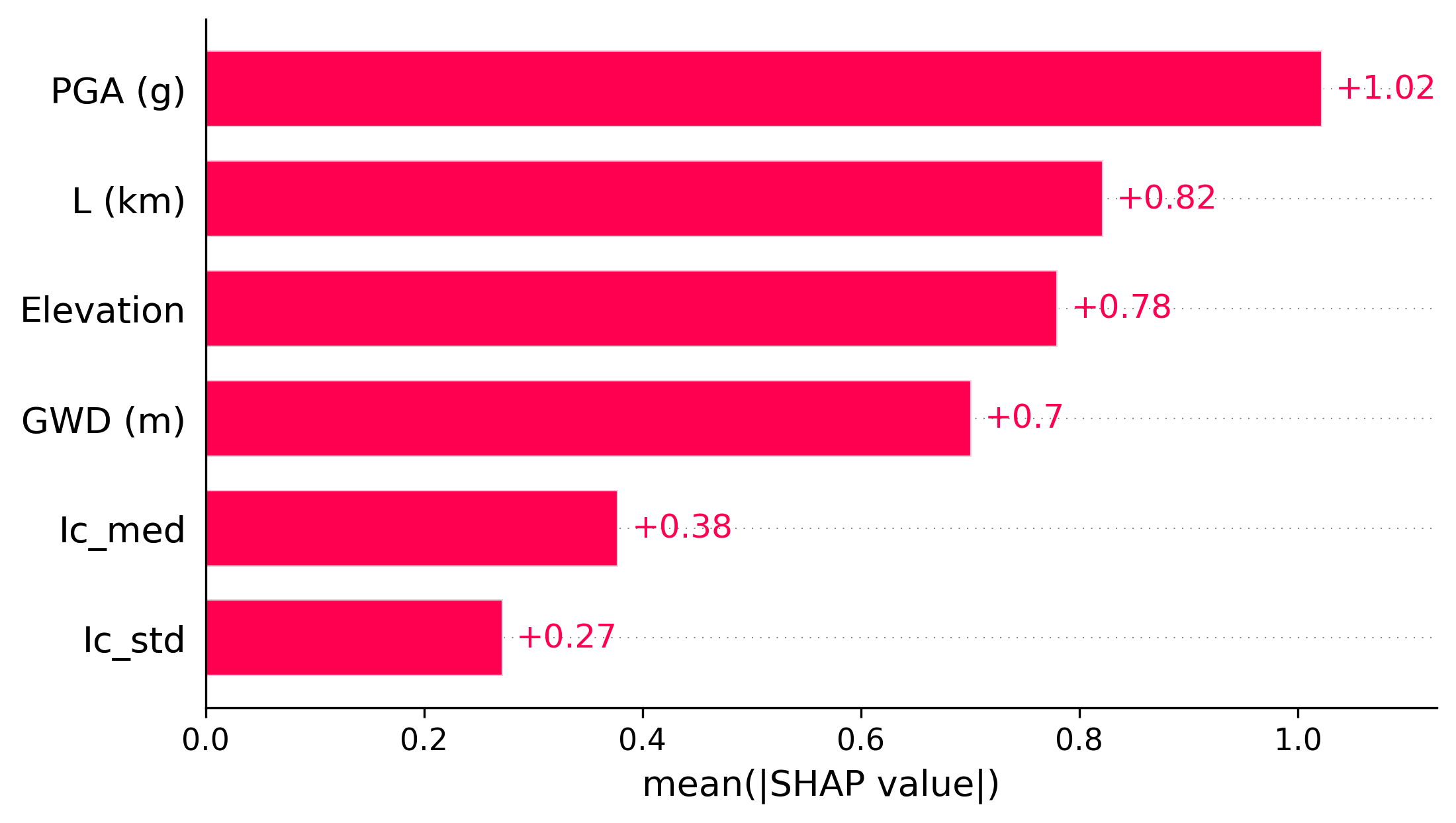}
        \caption{Feature importance}
        \label{fig:shap_bar_modelc}
    \end{minipage}
\setcounter{figure}{15}
\setcounter{subfigure}{2}
    \begin{minipage}[b]{0.45\textwidth}
        \includegraphics[width=\linewidth]{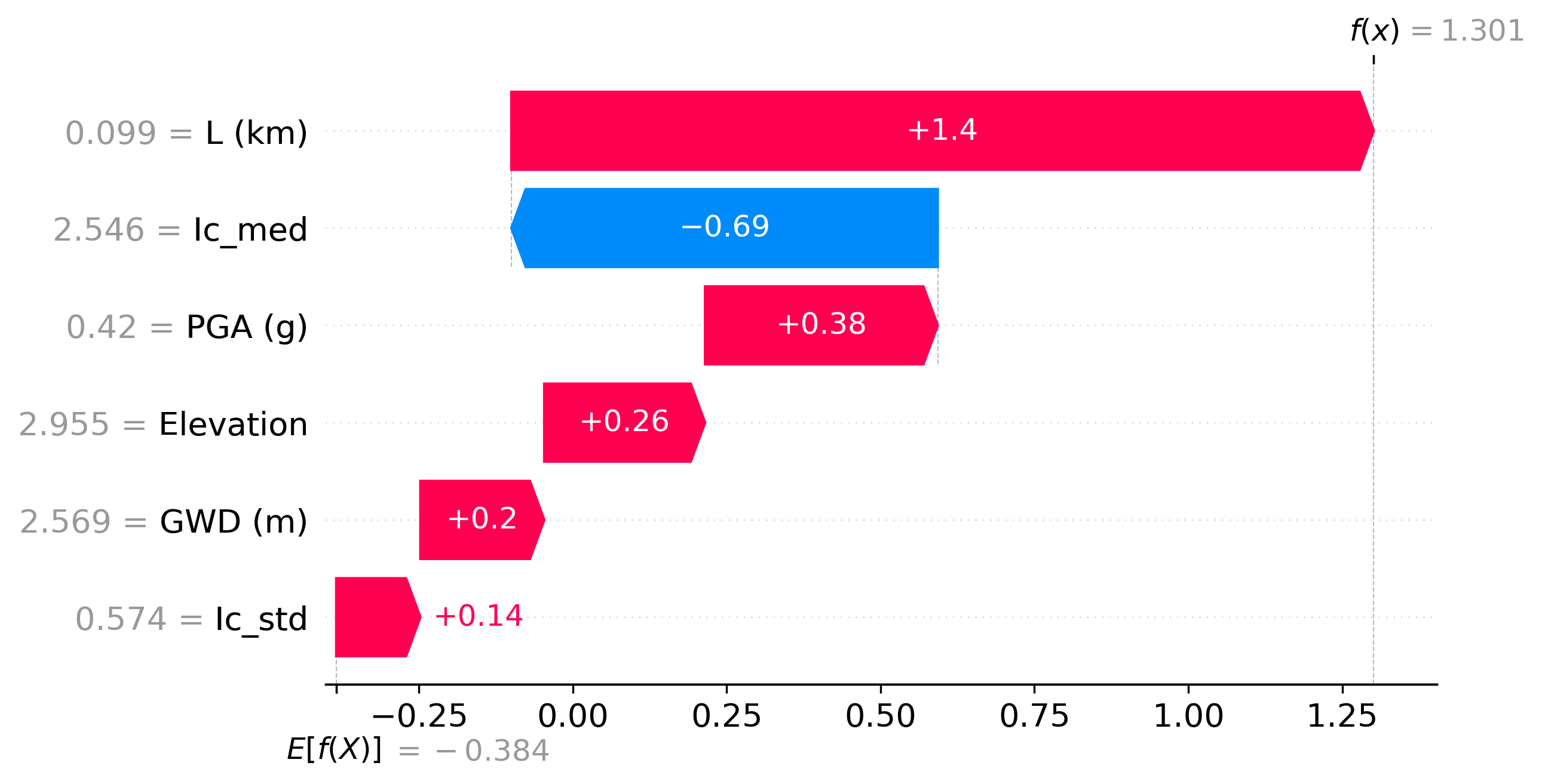}
        \caption{Site 5252 in Model C (True Positive)}
        \label{fig:c_5252}
    \end{minipage}
\setcounter{figure}{15}
\setcounter{subfigure}{3}
    \begin{minipage}[b]{0.45\textwidth}
        \includegraphics[width=\linewidth]{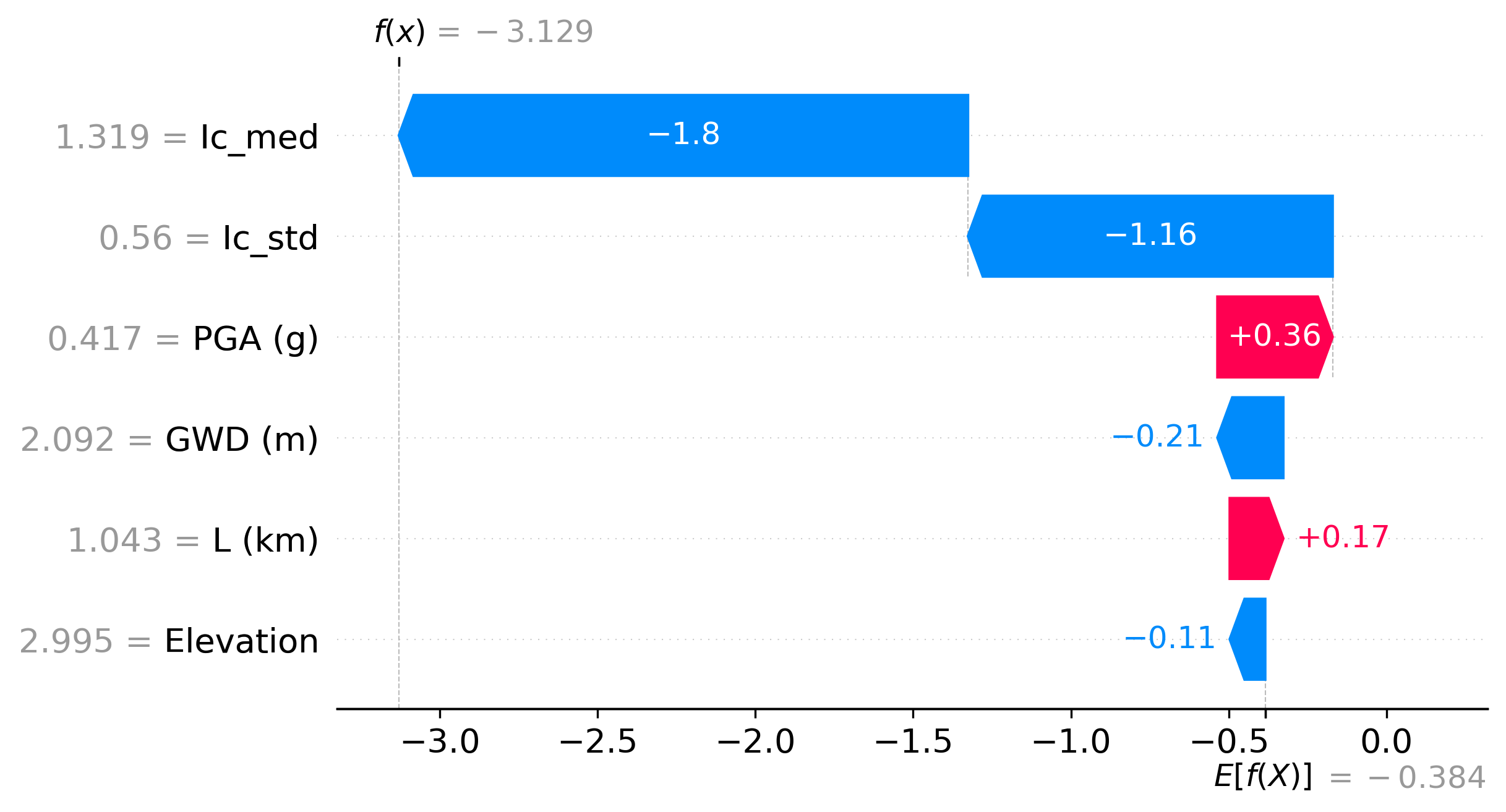}
        \caption{Site 17049 in Model C (True negative)}
        \label{fig:c_17049}
    \end{minipage}
\setcounter{figure}{15}
\setcounter{subfigure}{-1}
    \caption{Global and local explanation of Model C
    \textbf{(a)} Global explanation;
    \textbf{(b)} Feature importance;
    \textbf{(c)} Site 5252 in Model C (True positive);
    \textbf{(d)} Site 17049 in Model C (True negative).
    }
    \label{fig:model_c}
\end{subfigure}

\begin{figure}
\begin{center}
\includegraphics[width=8cm]{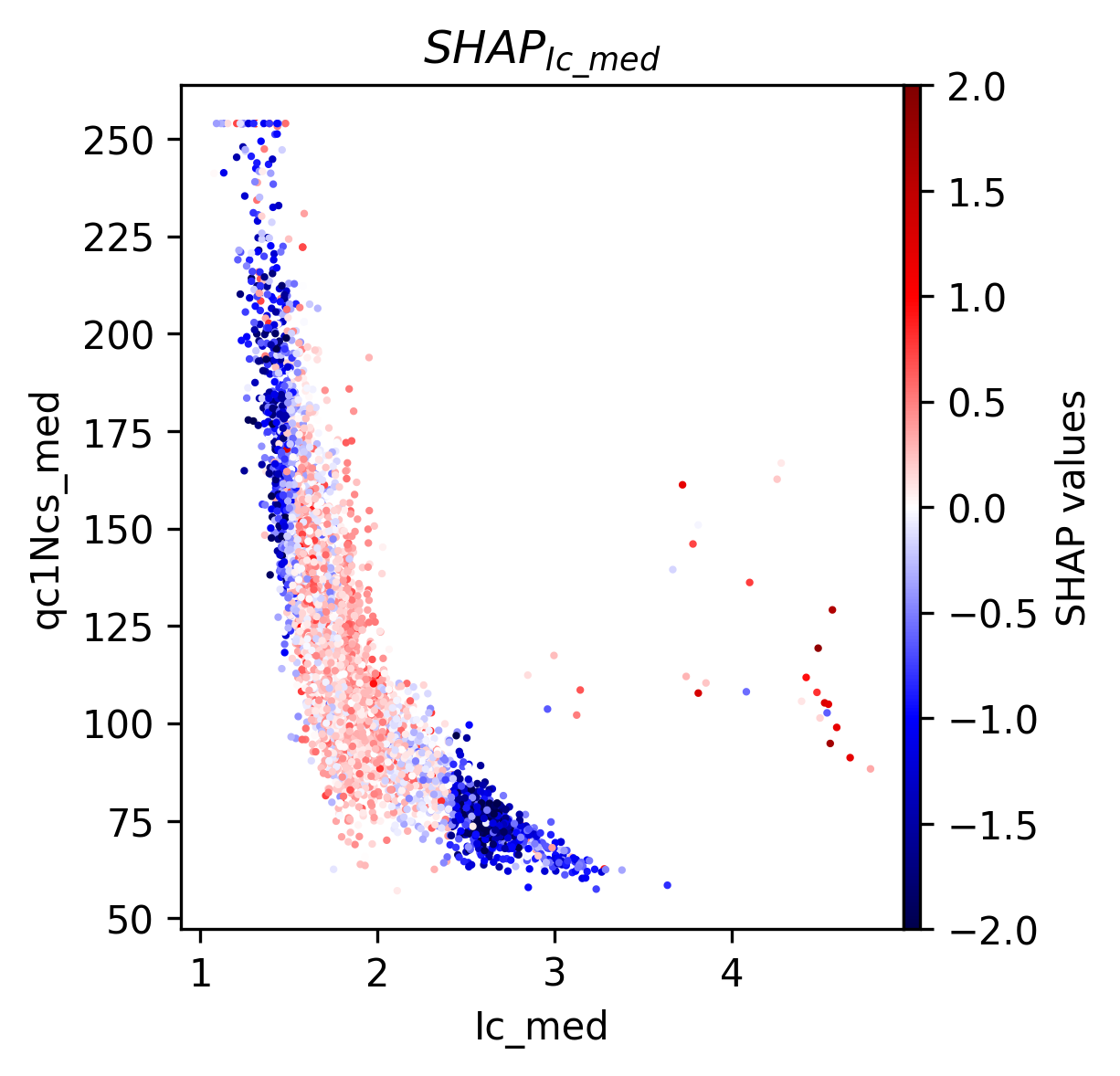}
\end{center}
\caption{SHAP values for medians of $I_c$ in Model C}
\label{fig:shap_ic_modelC}
\end{figure}

In~\Cref{fig:shap_summary_modelc,fig:shap_bar_modelc}, we present the global explanation for Model C. The results show that the most important feature is PGA, followed by L, elevation, and GWD. Similar to the ranking in~\Cref{fig:model_b_global}, the CPT features, the median and standard deviation of $I_{c}$ are the least important. To further examine Model C's behavior, we look at the local explanations for Site 5252 and Site 17049 (see~\Cref{fig:c_5252,fig:c_17049}). 

For Site 5252, the vicinity to the river (L = 99m) significantly impacts the prediction, with a SHAP value of 1.4. However, the median $I_c$ value of 2.546 has a negative contribution with a SHAP value of -0.69. Despite this, the effect of proximity to the river outweighs the $I_c$ feature, leading to a true positive prediction, which was incorrectly classified as a false negative by Model C.

In the case of Site 17049, Model C has learned that a low $I_c$ implies a high $q_{c1Ncs}$ and thus, low likelihood of lateral spreading. This is reflected in the decisive SHAP value of -1.8 for the $I_c$ feature, resulting in a true negative prediction. This suggests that the high correlation between median $I_c$ and median $q_{c1Ncs}$ allows Model C to  learn the underlying physics, as shown in  ~\Cref{fig:shap_ic_modelC}, using $I_c$ only.

We aimed to avoid feature coherence and overfitting by minimizing the number of input features, developing simpler and more explainable models. However, this approach may make Model C less generalizable to other sites, as its performance relies on the negative correlation between $q_{c1Ncs}$ and $I_c$ observed in the training dataset.

\section{Conclusion}

Our investigation into using SHapley Additive exPlanations (SHAP) for interpretability in machine learning models targeting lateral spreading predictions highlights four major insights. First, SHAP's application clarifies the XGB model's decision-making process, enhancing model transparency and fostering informed decision-making. Second, we identified incorrect learning patterns within the model, especially regarding PGA's impact, underscoring the need to align model predictions with domain knowledge. Third, incorporating CPT data into the XGB model, although did not change the predictive accuracy, improved model explanations. Notably, the model identified negative SHAP values for firm sandy and clayey soil, aligning with our understanding of soil types and liquefaction behavior. The XGB model accurately learned the relationship between soil characteristics inferred from CPT data and the likelihood of lateral spreading events conforming to the domain knowledge. Lastly, the study emphasizes the need for model optimization to enhance efficiency and generalizability. This research underlines the critical role of explainable AI in improving the reliability and utility of ML models for geotechnical engineering, contributing to more accurate hazard assessments and infrastructure protection.

\section*{Conflict of Interest Statement}

The authors declare that the research was conducted without any commercial or financial relationships that could be construed as a potential conflict of interest.

\section*{Author Contributions}

CH: Writing – original draft, Methodology, Formal analysis; KK: Writing – review \& editing, Conceptualization, Supervision, Funding acquisition; ER: Writing – review \& editing, Supervision.

\section*{Acknowledgments}
The authors thank the support of NHERI DesignSafe-CI for computing resources and data storage~\cite{Rathje2017DesignSafe}. 

\section*{Data Availability Statement}
The SHAP code is available under the MIT license on GitHub (\url{https://github.com/geoelements/xai-lateral-spreading}). The training dataset and trained models are published under a CC-By license on DesignSafe Data Depot~\citep{durante-2022}.

\bibliographystyle{Frontiers-Harvard} 
\bibliography{references}



\end{document}